\documentclass[preprint,3p,times,onecolumn]{elsarticle}
\usepackage[american]{babel}
\usepackage[font=normal]{caption}
\usepackage[font=normal,position=top]{subcaption}
\usepackage[tensorialbold]{userCommands}

\usepackage{tikz} 
\usetikzlibrary{calc,trees,positioning,arrows,chains,shapes.geometric,decorations.pathreplacing,decorations.pathmorphing,shapes,
    matrix,shapes.symbols,shapes.multipart,patterns,shapes,pgfplots.groupplots,spy,backgrounds}
\usetikzlibrary{spy,backgrounds}
\usetikzlibrary{decorations.markings}
\usetikzlibrary{arrows.meta}

\usepackage{pgfplots} 
\pgfplotsset{
  compat=newest,
  grid=both,
  tick label style={font=\normalsize},
  label style={font=\normalsize},
  legend style={font=\footnotesize},
  legend cell align={left},
  yticklabel style={/pgf/number format/fixed},
  colormap={tol}{[1cm] rgb255(0cm)=(120,28,129) rgb255(1cm)=(63,96,174) rgb255(2cm)=(83,158,182) rgb255(3cm)=(109,179,136) rgb255(4cm)=(202,184,67) rgb255(5cm)=(231,133,50) rgb255(6cm)=(217,33,32)},
  colormap={tol2}{[1cm] rgb255(0cm)=(217,33,32) rgb255(1cm)=(231,133,50) rgb255(2cm)=(202,184,67) rgb255(3cm)=(109,179,136) rgb255(4cm)=(83,158,182) rgb255(5cm)=(63,96,174) rgb255(6cm)=(120,28,129)}
}

\usepackage{array}
\newcolumntype{M}[1]{>{\centering\arraybackslash}m{#1}}
\newcolumntype{N}{@{}m{0pt}@{}}

\usepackage{amssymb}
\usepackage{lineno}

\usepackage{float}

\makeatother
\definecolor{Purple}{RGB}{120,28,129}
\definecolor{Orange}{RGB}{231,133,50}
\definecolor{Blue}{RGB}{63,96,174}
\definecolor{Red}{RGB}{217,33,32}
\definecolor{Duck}{RGB}{83,158,182}
\definecolor{Green}{RGB}{109,179,136}
\definecolor{Yellow}{RGB}{202,184,67}
\usepackage[normalem]{ulem}

\newcommand{\review}[1]{#1}

\newtheorem{remark}{Remark}

\usepackage{mathtools}

\begin{document}

\title{Multi-parameter optimization of attenuation data for characterizing grain size distributions and application to bimodal microstructures}


\author[mymainaddress]{Adrien Renaud\fnref{myfootnote}}
\fntext[myfootnote]{Current address: CEA -- Service de Recherches M{\'e}tallurgiques Appliqu{\'e}es, Universit{\'e} Paris-Saclay, 91191 Gif-sur-Yvette, France, e-mail: adrien.renaud2@cea.fr } 

\author[mymainaddress]{Bing Tie\corref{mycorrespondingauthor}}
\cortext[mycorrespondingauthor]{Corresponding author}
\ead{bing.tie@centralesupelec.fr}
\author[mymainaddress]{Anne-Sophie Mouronval}
\author[mymainaddress]{Jean-Hubert Schmitt}

\address[mymainaddress]{Universit\'e Paris-Saclay, CentraleSup\'elec, CNRS, Laboratoire de M\'ecanique des Sols, Structures et Mat\'eriaux, 91190 Gif-sur-Yvette, France}





\begin{abstract}
  In this paper, the effect on the ultrasonic attenuation of the grain size heterogeneity in polycrystals is analyzed.
  First, new analytical developments allowing the extension of the unified theory of \textsc{Stanke} and \textsc{Kino} to general grain size distributions are presented.
  It is then shown that one can additively decompose the attenuation coefficient provided that groups of grains are defined.
  Second, the study is specialized to a bimodal distribution of the grain size for which microstructures are numerically modeled by means of the software \textsc{Neper}.
  The additive partition of the attenuation coefficient into contributions coming from large and small grains motivates the derivation of an optimization procedure for characterizing the grain size distribution.
  The aforementioned approach, which is based on a least squares minimization, is at last presented and illustrated on both analytical and numerical attenuation data.
  It is thus shown that the method provides satisfying approximations of volume fractions of large grains and modal equivalent diameters from the frequency-dependent attenuation coefficient.

\end{abstract}

\maketitle
\section{Introduction}
\label{sec:introduction}
In the context of non-destructive evaluation of polycrystals, the attenuation resulting from the scattering of ultrasonic waves due to the grain microstructure can be used to deduce some features of the latter.
Indeed, it has first been emphasized that the scalar quantity $\alpha$, called the attenuation coefficient, can be assumed proportional to $d^{n-1}f^n$, where $d$ quantifies the mean grain size and $f$ is the frequency of the ultrasonic signal \cite{smith,Nicoletti}.
The value of the exponent $n$ changes for the three scattering regions, which can be distinguished by the ratio between the wavelength $\lambda$ and the mean grain size $d$ (\textit{i.e.} the Rayleigh, stochastic and \review{geometric} scattering domains for $\lambda \gg d$, $\lambda \approx d$ and $\lambda \ll d$ respectively).
Notice, however, that the limits between the three domains are not clearly specified and seem to be material-dependent \cite{botvina}.
Thus, the relation between the morphology, namely, the size and shape of grains, and the amplitude decay of elastic waves propagating in a polycrystal has been widely exploited for the non-destructive evaluation of the grain size as in, among others, references \cite{PALANICHAMY,botvina,sundin,bouda,li,levesque,sarkar,garcin,keyvani}.

Regarding theoretical contributions, two seminal theories can be cited.
On the one hand, \textsc{Stanke} and \textsc{Kino} proposed a unified theory valid in all frequency regions. 
These developments were based on the second-order Keller approximation for weak anisotropic media composed of spherical grains with the same size and assuming single-scattering \cite{Stanke_Kino84,Keller64}.
The weak-scattering Born approximation was further used to derive an explicit formula for the attenuation coefficient in single-phase and untextured three-dimensional polycrystals with a cubic symmetry.
On the other hand, \textsc{Weaver} developed a multiple-scattering formalism for the mean Green\review{'s} function and the covariance of the Green\review{'s} function.
The latter, an energy density, is found to obey a radiative transfer equation for which a diffusion limit can be taken \cite{Weaver_modeConversion}.
By invoking the Born approximation, closed forms of the attenuation coefficient were obtained, which give good comparisons with \textsc{Stanke} and \textsc{Kino}'s model for both Rayleigh and stochastic regions but fail in the \review{geometric} region \cite{Kube_JASA2017}.
Recently, using the framework proposed by \textsc{Stanke} and \textsc{Kino}, explicit formulas of the attenuation coefficient in both two and three-dimensional cases were developed for untextured polycrystals with equiaxed grains with cubic symmetry \cite{Xue_2Dalpha}.
A rigorous analysis of the dimensionality of the grain scattering-induced attenuation within those media was then carried out.
The latter theoretical framework is considered and extended in this paper.

\review{
  In practice, the use of such analytical formulas in test setups is not straightforward. 
  First, ultrasounds of a broad frequency range can be used and involve more than one scattering domain for a given grain size.
  Hence, to interpret the experimental data, analyses are generally carried out by identifying, in a more or less ad hoc way, an apparent value of the exponent $n$.
  For example, in \cite{garcin}, the experimental analysis is performed with an apparent value $n = 3$.} 
\review{Second}, the grain morphology can itself complicate non-destructive testing.
For instance, it has been shown that the width of the grain size distribution strongly affects the attenuation curve except when the scattering region is restricted to the Rayleigh region for the whole distribution \cite{smith,Nicoletti,Arguelles2017}.
This makes difficult the determination of the grain size from the ultrasonic attenuation without any coupled metallographic observation \cite{nicoletti1997}.
In addition, polycrystals that underwent recrystallization or heterogeneous grain growth, leading to a bimodal distribution of the grain size \cite{Boyce2015_bimodal}, constitute another difficulty for accurately determining the mean grain size and hence, its evolution.

The present work focuses on the limitation of ultrasonic testing for polycrystals associated with a bimodal distribution of the grain size.
In the following, microstructures exhibiting such statistics are referred to as \textit{bimodal} ones for simplicity.
Combining theoretical and numerical approaches, this work first aims at identifying the effect of a bimodal distribution of the grain size on the scattering-induced attenuation.
This aspect requires the ability to numerically model bimodal microstructures with a realistic grain shape and to take into account the corresponding morphology to compute the attenuation coefficient.
Following former works \cite{Liu2008,Ryzy2018,Pamel2018}, this can be done through the determination of the spatial correlation function (also called two-point correlation or autocorrelation function).
Second, the identified attenuation response is used to develop a characterization procedure for bimodal microstructures based on reference attenuation results that can come from analytical, numerical or experimental data.
This work then represents a first step towards a real-time monitoring of polycrystalline materials, for which a bimodal distribution of the grain size can be desired to achieve some chemical or mechanical properties \cite{Chakrabarti2009_bimodal,Sabzi2016_bimodal,Azizi2007_bimodal_biphase,Mahesh2012bimodal}. 

The comparison of experimental data resulting from multiple-scattering with analytical results based on single-scattering only may lead to significant gaps \cite{Zhang04,zeng}.
The gaps between analytical and experimental attenuation relating to taking into account multiple-scattering, which make difficult the \textit{in-situ} evaluation of the grain size, can be circumvented by resorting to finite element modeling.
Indeed, this allows to take into account the grain structures without using simplifying assumptions.
More specifically, discontinuous Galerkin methods, based on a piece-wise polynomial approximation of the solutions of hyperbolic problems \cite{Cockburn,TIE2018}, \review{are} considered \review{in the present work}.
This class of methods enables, through the solution of Riemann problems at the element interfaces, to grasp the complex physical process of wave propagation in polycrystals.
%
Therefore, numerical approaches allow, by considering realistic morphologies of the grains resulting from EBSD or numerical generation as proposed by the software \textsc{Neper} \cite{Neper}: (i) accounting for multiple-scattering coming from the complex reflection of waves at grain interfaces; (ii) computing a full-field solution for any sample geometries.

The paper is organized as follows.
First, Section \ref{sec:theory} is devoted to theoretical aspects to give a coherent framework.
The attenuation coefficient formulas derived in \cite{Xue_2Dalpha} for two and three-dimensional problems are recalled and extended to a generic two-point correlation function.
It is then shown that for bimodal microstructures made of equiaxed grains, this function leads to an additive decomposition of the attenuation coefficient.
Second, a procedure to numerically model two and three-dimensional polycrystals with a bimodal distribution of the grain size and a  realistic morphology using the software \textsc{Neper} is presented in Section \ref{sec:microstructure}.
The analysis of those bimodal microstructures shows that the numerical models satisfy the additive property of the spatial correlation function.
Then the frequency-dependent attenuation coefficient is analytically evaluated and the influence of a bimodal distribution of the grain size is emphasized in Section \ref{sec:procedure}.
The inverse characterization procedure based on these observations is at last presented and illustrated with analytical attenuation results, which are supplemented with numerical data in Section \ref{sec:numerical_results}.



\section{Theoretical considerations}
\label{sec:theory}
We first propose to briefly recall the derivation of attenuation coefficient formulas provided by the unified theory of \textsc{Stanke} and \textsc{Kino} \cite{Stanke_Kino84}.
These developments involve the spatial correlation function that accounts for the microstructural morphology of polycrystalline media and is therefore of major importance for the computation of attenuation curves.
As a first result of this paper, semi-explicit equations are derived for the calculation of the frequency-dependent attenuation coefficient accounting for a sample-defined autocorrelation function in two and three space dimensions.
Attention is next paid to the two-point correlation function for polycrystals whose grain size follows a bimodal distribution.
It is then shown that in such cases the spatial correlation function, and hence the frequency-dependent attenuation coefficient, breaks down additively into two contributions.

\subsection{Analytical modeling of the attenuation}
\label{sec:alphaTH}

Let $\Omega \in \Rbb^{\text{dim}}$ ($\text{dim}=2,3$) be a domain occupied by a polycrystalline material characterized by the position-dependent elastic stiffness tensor $\Cbb(\vect{x})$ and the constant mass density $\rho$.
Consider now the time harmonic elastic wave equation in $\Omega$ with no source term:
\begin{equation}
  \label{eq:balanceLaw_Fourier}
  \vect{L}\(\vect{u}(\vect{x},\omega)\) \equiv \rho \omega^2 \vect{u}(\vect{x},\omega) + \nablat \cdot \( \Cbb(\vect{x}) : \tens{\eps}(\vect{u}(\vect{x},\omega)) \) = \vect{0}
\end{equation}
where $\vect{u}(\vect{x},\omega)$ is the Fourier transform with respect to time of the displacement field $\vect{u}(\vect{x},t)$, $\nablat \cdot (\bullet)$ is the divergence operator, and $\tens{\eps}(\bullet)$ the linearized strain tensor.
The domain $\Omega$ is composed of $N^g$ non-overlapping subdomains $\Omega^i$, referred to as grains, such that:
\begin{equation}
  \label{eq:subdomains}
  \Omega = \bigcup\limits_{i=1}^{N^g} \Omega^i
\end{equation}

Assuming \textit{single-phase}, \textit{untextured} and \textit{weakly-scattering} polycrystals, explicit formulas of attenuation can be derived for problems in two and three space dimensions \cite{Xue_2Dalpha} by using the unified theory of \textsc{Stanke} and \textsc{Kino} \cite{Stanke_Kino84} based on a general formulation established by \textsc{Karal} and \textsc{Keller} \cite{Keller64}.
These developments are constructed upon the search for the expected wave solution of an ensemble of possible inhomogeneous media $\left\langle \vect{u}(\vect{x},\omega)\right\rangle$, where $\left\langle \bullet \right\rangle$ denotes an ensemble averaging operator.
For a particular heterogeneous medium, the inhomogeneity degree can be quantified by the deviation of its elastic tensor compared to that of an equivalent homogeneous medium: $\delta \Cbb(\vect{x})= \Cbb(\vect{x}) - \Cbb^0$, $\Cbb^0=\left\langle \Cbb \right\rangle_\Theta$ being the Voigt average over all crystallographic orientations $\Theta$ \review{as originally chosen in Stanke and Kino's or Weaver's works} \cite{Stanke_Kino84,Weaver_modeConversion}.
\review{However,  it worth noticing that this choice is not unique. For example, the Reuss average or the self-consistent (SC) average can be used \cite{Kube_WM2015}, but are not considered here.}
From now on, ``$0$'' superscripts refer to the equivalent homogeneous medium.

Using the second-order Keller approximation, the following explicit equation for $\left\langle \vect{u}(\vect{x},\omega) \right\rangle$ can be obtained:
\begin{equation}
  \label{eq:keller}
  \vect{L}^0\(\left\langle \vect{u}(\vect{x},\omega) \right\rangle\) - \left\langle \vect{L}^1\(\int_\Omega \tens{G}^T(\vect{x}',\vect{x})\) \cdot \vect{L}^1(\left\langle \vect{u}(\vect{x}',\omega)  \right\rangle ) d\vect{x}'\right\rangle = 0
\end{equation}
in which $\tens{G}$ is the dyadic Green\review{'s} function tensor and $\vect{L}^0(\bullet)$ and $\vect{L}^1(\bullet)$ are respectively the homogeneous and perturbation operators defined as:
\begin{align}
  \label{eq:Loperators}
  & \vect{L}^0\(\vect{u}(\vect{x},\omega)\) = \rho \omega^2 \vect{u}(\vect{x},\omega) + \nablat \cdot \( \Cbb^0 : \tens{\eps}(\vect{u}(\vect{x},\omega)) \) \\
  & \vect{L}^1\(\vect{u}(\vect{x},\omega)\) = \nablat \cdot \( \delta \Cbb : \tens{\eps}(\vect{u}(\vect{x},\omega)) \)
\end{align}

The \review{following assumptions:}
\begin{itemize}
\item the single-phase setup along with the fact that the deviation of the elastic tensor in each grain is constant and written as $\delta \Cbb^g$,
\item \review{the elastic tensor components and the geometric characteristic functions of the grains vary independently,}
\item \review{the deviation in the elastic tensor components vary independently from grain to grain,}
\end{itemize}
allow to rewrite the autocorrelation function of the elastic tensor as:
\begin{equation}
  \label{eq:spatial_autocorrelationCW}
  \left\langle \delta \Cbb (\vect{x}) \otimes \delta \Cbb (\vect{x}') \right \rangle = \left\langle \delta \Cbb^g \otimes \delta \Cbb^g) \right \rangle_\Theta W(\vect{r})
\end{equation}
where $\vect{r}=\vect{x}-\vect{x}'$, and $W(\vect{r})$ is the spatial correlation function of two points $\vect{x}$ and $\vect{x}'$.
For the untextured media under consideration (\textit{i.e.} statistically isotropic), a spherical symmetry is assumed: $W(r)\equiv W(r\vect{n}), \forall \vect{n}$ with $\norm{\vect{n}} = 1$ \cite{Torquato_book}.

By seeking plane wave solutions of equation \eqref{eq:keller} and using the Born approximation (see \cite{Xue_2Dalpha} for an exhaustive derivation), longitudinal and transverse attenuation coefficients can be written.
Both expressions are composed of one contribution induced by scattering into a same type of wave and another one generated by mode conversion \cite{Weaver_modeConversion}:
\begin{equation}
  \label{eq:alpha_decomp}
  \alpha^L = \alpha^{LL} +\alpha^{LT} \quad ; \quad \alpha^T = \alpha^{TT} +\alpha^{TL},
\end{equation}
with, in direction $\vect{e}_J$: 
\begin{equation}
  \label{eq:alpha_integral}
  \alpha^{\beta \gamma} = \sum_{k,l,m,n}\text{Im}\[ \frac{k_{0\beta}\left\langle \delta C_{J^\beta J kl} \delta C_{mn J^\beta J} \right\rangle}{2C^0_{J^\beta J J^\beta J}} \int_\Omega G^\gamma_{km} D^{\beta,J}_{ln} d\Omega\] 
\end{equation}
In equation \eqref{eq:alpha_integral}, Im$\[\bullet\]$ denotes the imaginary part, $k_{0\beta}$ is the wave constant of a $\beta$-type wave in the Voigt average homogeneous reference medium and $\vect{e}_{J^\beta}$ indicates the polarization direction of the $\beta$-wave that propagates in the direction $\vect{e}_J$.
Namely, $J^\beta = J\delta_{ \beta L} + P \delta_{ \beta T}$, with $\vect{e}_P \perp \vect{e}_J$ for a transverse wave. 
Moreover, the components of the Green\review{'s} tensor $\tens{G^\gamma}$ and the expression of $\tens{D}^{\beta,J}(\vect{r})$ are respectively \cite{Xue_2Dalpha}:
\begin{align}
  \label{eq:Ggamma_expression}
  & \tens{G}^\gamma(\vect{r})= \frac{1-c(\gamma)}{4\pi \rho \omega^2}\( A_{rr}(k_{0\gamma}r)\frac{\vect{r}\otimes\vect{r}}{r^2} -A_{I}(k_{0\gamma}r)\tens{I}\)\\
  \label{eq:Dbeta_expression}
  &\tens{D}^{\beta,J}(\vect{r})= \nablat_{\vect{r}}\(\nablat_{\vect{r}}\(W(r) e^{ik_{0\beta}\vect{e}_{J} \cdot \vect{r}}\)\)
\end{align}
where $r=\norm{\vect{r}}$, $c(\gamma)$ is a constant equal to $0$ for $\gamma=L$ and $2$ for $\gamma=T$, $i=\sqrt{-1}$, and the expression of the functions $A_{rr}(\bullet)$ and $A_I(\bullet)$ depending on the space dimension can be found in \ref{sec:alpha_formula}.

The spatial correlation function $W(r)$ describes the possibility that two points lie in the same crystal and therefore accounts for the morphology of the polycrystalline microstructure.
A semi-analytical attenuation model based on a \textit{sample-defined} spatial correlation function \cite{Ryzy2018} has been employed to compute attenuation coefficient curves that are better suited to the grain morphology of three-dimensional microstructures.
To make a similar model available for both two and three-dimensional solids, the first contribution of this work consists in extending the analytical results presented in \cite{Xue_2Dalpha} by expanding generically the equation  \eqref{eq:Dbeta_expression} as follows:
\begin{equation}
  \label{eq:generic_D}
  \begin{split}
    \tens{D}^{\beta,J}(\vect{r})= e^{ik_{0\beta}\vect{e}_{J} \cdot \vect{r}}&\[ \( W''(r) - \frac{W'(r)}{r}\)\frac{\vect{r}\otimes\vect{r}}{r^2} -k_{0\beta}^2 W(r)  \vect{e}_{J}\otimes\vect{e}_{J} \right. 
    \left.+ \frac{W'(r)}{r}\{ \tens{I} + ik_{0\beta} \( \vect{r} \otimes \vect{e}_{J} + \vect{e}_{J} \otimes \vect{r} \) \}\]  
  \end{split}
\end{equation}
in which usual prime notations are used for the derivatives of $W(r)$.
From equation \eqref{eq:generic_D}, it is possible to derive semi-analytical models for two and three-dimensional problems that enable the computation of attenuation coefficient curves based on \textit{sample-defined} spatial correlation functions (see \ref{sec:alpha_formula} for details).

\subsection{The spatial correlation function} 
\label{sec:Wfunc}

Let us recall one important feature of the spatial correlation function $W(r)$ in polycrystalline materials \cite{Torquato_book} that allows rewriting the attenuation coefficient in particular cases.
In analogy with multi-phase materials \cite{Torquato_book,Arguelles2017}, the probability that the line segment of length $r$ lies entirely, when thrown randomly in $\Omega$, in one grain belonging to some family $k$ occupying the domain $\Omega^k$, is quantified by the autocorrelation function $W_{\Omega}^k(r)$:
\begin{equation}
  \label{eq:W_mono}
  W_{\Omega}^{k}(r) =  F_k \frac{\int_{y=0}^\infty (y-r) p^{k}(y) H(y-r) dy}{\int_{y=0}^\infty y p^k(y) dy} = F_k W_{\Omega^k}^{k}(r)
\end{equation}
In equation \eqref{eq:W_mono}, $H(\bullet)=\frac{\abs{\bullet}+\bullet}{2\bullet}$ is the Heaviside step function and $p^k(y)$ is the chord-length probability density function of the family $k$.
Furthermore, $F_k$ is the volume fraction of the phase $k$ in $\Omega$ and $W_{\Omega^k}^{k}(r)$ represents the probability that the line segment is contained in on grain when thrown randomly in $\Omega^k$.


The spatial correlation function in the whole domain $\Omega$, which corresponds to a sum of probabilities, therefore reads:
\begin{equation}
  \label{eq:W_multiphase}
  W(r) 
  = \sum_{k} F_k W_{\Omega^k}^{k}(r) 
\end{equation}

By denoting $\alpha^{\beta\gamma}_{k}$ the attenuation coefficient of the single-phase material $k$, the above decomposition, once introduced in equation \eqref{eq:Dbeta_expression}, yields the following additive decomposition of the attenuation coefficient in the multi-phase case:
\begin{equation}
  \label{eq:multiphase_decomposition}
  \alpha^{\beta\gamma} = \sum_{k} F_{k}\alpha^{\beta\gamma}_{k} 
\end{equation}
This attenuation series can be seen as a discrete version of the formula used in \cite{smith,Nicoletti,nicoletti1997}, which has been developed based on \textsc{Roney}'s model \cite{Roney}. 

The discussion is now specialized to microstructures whose distribution of the grain size is bimodal.
In this case, the grains can be classified with respect to the family they belong to, that is, the one of small grains (SG) or the one of large grains (LG).
It then appears that, in analogy with the multi-phase case, the spatial correlation function in bimodal microstructures can be written as:
\begin{equation}
  \label{eq:W_bimodal}
  W(r) = F_{LG}W^{LG}(r)  + F_{SG}W^{SG}(r) 
\end{equation}
and the attenuation coefficient takes the form of a convex combination:
\begin{equation}
  \label{eq:bimodal_decomposition}
  \alpha^{\beta\gamma}_{\text{Bimodal}} = F_{LG}\alpha^{\beta\gamma}_{LG} + (1-F_{LG})\alpha^{\beta\gamma}_{SG}
\end{equation}

Equation \eqref{eq:bimodal_decomposition} highlights that in addition to the dependence of the attenuation coefficient $\alpha^{\beta\gamma}$ on the frequency and the mean grain size, the volume fraction plays an important role for bimodal microstructures.
Hence, given the analytical formula \eqref{eq:alpha_integral}, a numerical procedure can be developed in order to define (i) the modal grain sizes of the distribution; (ii) the volume fraction of each family.
This is the object of section \ref{sec:procedure}.
\section{Microstructures modeling}
\label{sec:microstructure}
A procedure to numerically model two and three-dimensional bimodal microstructures is presented in this section.
Examples of geometry resulting from this approach are shown as well as the distribution of the equivalent diameter in the samples.
In addition, the sample-defined autocorrelation functions computed in these polycrystals are presented and analyzed with regard to the previously highlighted additive decomposition property.

\subsection{Generation of the geometry}
\label{sec:generation-geometry}

The grains geometry associated with the microstructures considered here are constructed by means of the software \textsc{Neper} \cite{Neper}.
Based on Laguerre tessellations combined with optimization processes, \textsc{Neper} allows generating microstructures whose morphological properties follow some statistical distribution \cite{Quey_gg}.
This approach is particularly interesting in order to model bimodal distributions of the grain size that could not result from classical Voronoi tessellations (which are a particular case of Laguerre tessellations).
The measure used to characterize the grain size is the equivalent diameter $d$, defined for two-dimensional and three-dimensional cases as the diameter of the circle of equivalent area and the sphere of equivalent volume respectively. 
The procedure is as follows:
\begin{itemize}
\item[1--] Given the two modal diameters $(d^{SG},d^{LG})$, the volume (\textit{resp.} the area) of the sample $V$ and the volume fraction $F_{LG}$, compute the number of spherical (\textit{resp.} circular) grains in each family: $n^{SG}$ and $n^{LG}$.
\item[2--] Compute the arithmetic average of the equivalent diameter as:
  \begin{equation}
    \bar{d}=\frac{n^{SG}d^{SG} + n^{LG}d^{LG}}{n^{G}},
  \end{equation}
  in which $n^G=n^{SG} + n^{LG}$ is the total number of grains.
  It is worth noticing that $\bar{d}$ does not correspond to the equivalent diameter but is only used as a parameter within \textsc{Neper}.
\item[3--] Construct a log-normal bimodal distribution of the equivalent diameter as the arithmetic average of the monomodal distributions:
  \begin{equation}
    d \sim \frac{n^{SG}}{n^{G}}\Lc_\Nc (\mu=\frac{d^{SG}}{\bar{d}},\sigma^{SG}) +  \frac{n^{LG}}{n^{G}}\Lc_\Nc (\mu=\frac{d^{LG}}{\bar{d}},\sigma^{LG})
  \end{equation}
  where $\mu$ denotes the expectation, and $\sigma^i$ and $\frac{n^{i}}{n^{G}}$ are the standard deviation and the numerical fraction of the grains of family $i$ respectively.
\end{itemize}
\begin{remark}
  The log-normal distribution is here chosen for the equivalent diameter as it is well representative of microstructures that underwent grain-growth for a standard deviation $\sigma=0.35$ \cite{Quey_gg}.
  In this paper, the standard deviations used are set to a low value $\sigma^{LG}=\sigma^{SG}=0.06$ in order to better distinguish two families for bimodal microstructures.
\end{remark}

\begin{figure}[!ht]
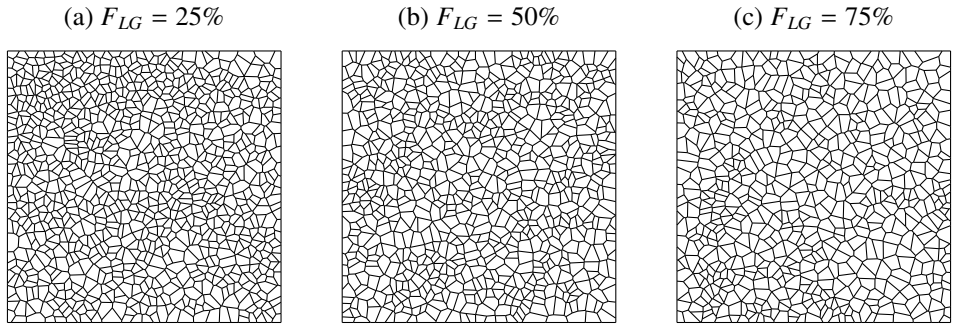

  \centering
  \subcaptionbox{$F_{LG}=25 \%$}{\input{pgfFigures/square240_160_FLG25.tex}}\qquad
  \subcaptionbox{$F_{LG}=50 \%$}{\input{pgfFigures/square240_160_FLG50.tex}}\qquad
  \subcaptionbox{$F_{LG}=75 \%$}{\input{pgfFigures/square240_160_FLG75.tex}}
  \caption{Two-dimensional bimodal microstructures with the modal equivalent diameters $d^{SG}=160\mu m$ and $d^{LG}=240\mu m$ for three different volume fractions of large grains.}
  \label{fig:2DbimodalExamples}
\end{figure}

\begin{figure}[!ht]
  \centering
  \subcaptionbox{$F_{LG}=25 \%$}{\includegraphics[trim=6cm 4.5cm 6cm 4.5cm, clip,width=0.3\linewidth,height=0.3\linewidth]{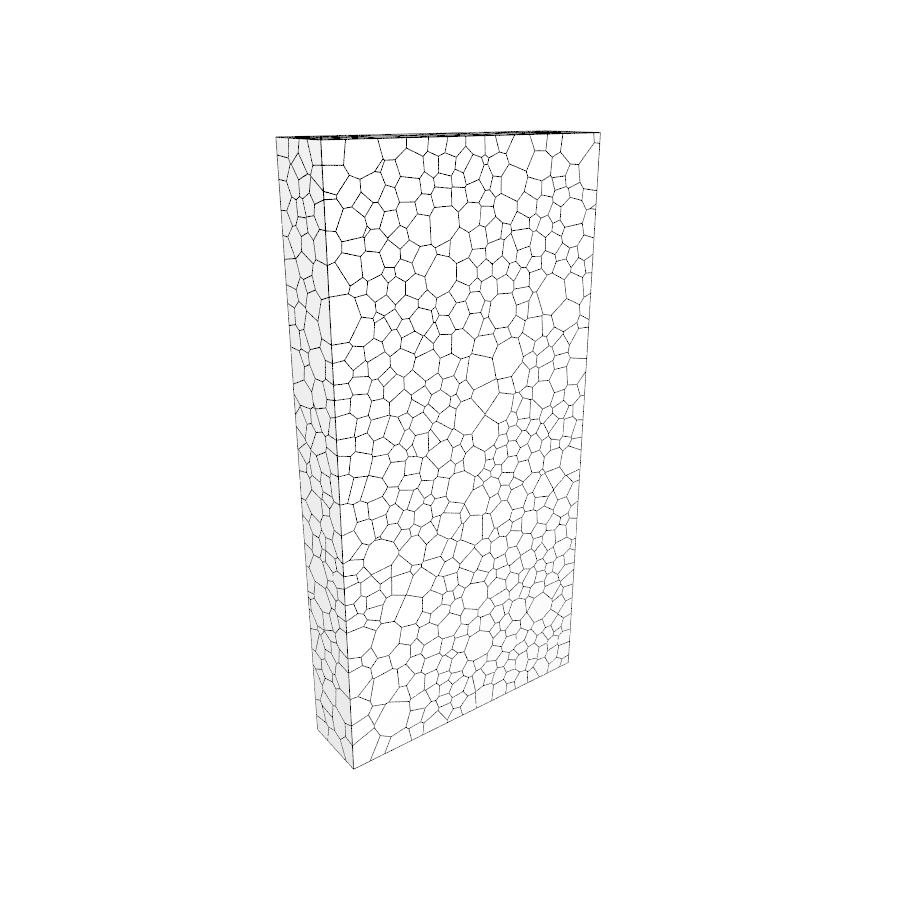}}
  \subcaptionbox{$F_{LG}=50 \%$}{\includegraphics[trim=6cm 4.5cm 6cm 4.5cm, clip,width=0.3\linewidth,height=0.3\linewidth]{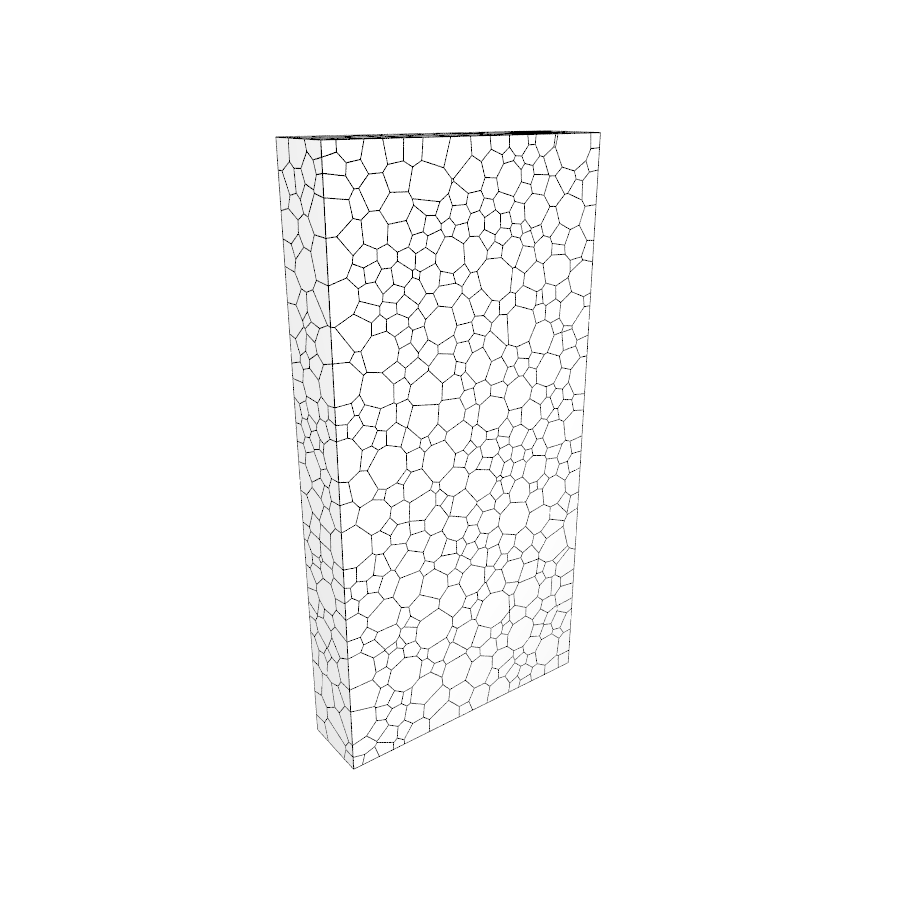}}
  \subcaptionbox{$F_{LG}=75 \%$}{\includegraphics[trim=6cm 4.5cm 6cm 4.5cm, clip,width=0.3\linewidth,height=0.3\linewidth]{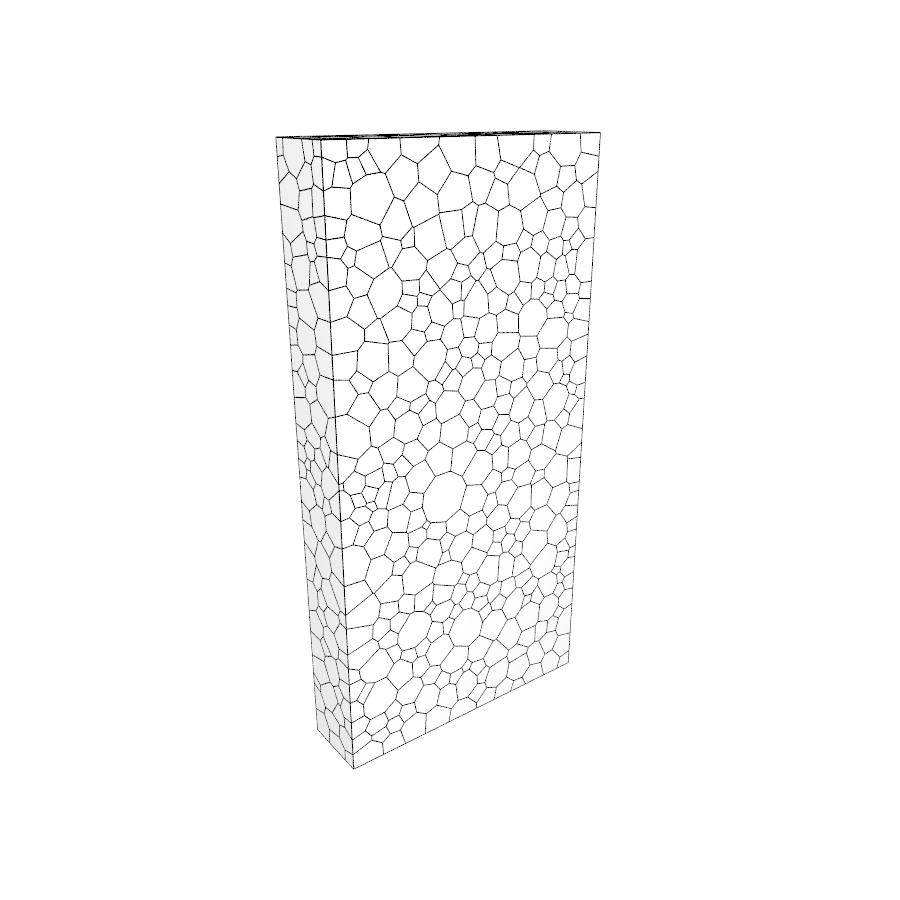}}
  \caption{Three-dimensional bimodal microstructures with the modal equivalent diameters $d^{SG}=160  \mu m$ and $d^{LG}=240  \mu m$ for three different volume fractions of large grains.}
  \label{fig:3DbimodalExamples}
\end{figure}

Examples of two and three-dimensional microstructures generated by following the above procedure are depicted in figures \ref{fig:2DbimodalExamples} and \ref{fig:3DbimodalExamples} respectively.
These geometries are obtained for the modal equivalent diameters of $d^{SG}=160 \mu m$ and $d^{LG}=240 \mu m$ and three different volume fractions of large grains.

In what follows, two and three-dimensional domains of dimensions $(x_1,x_2)\in \[0,9.6\]\times \[0,4.8\]$mm$^2$ and $(x_1,x_2,x_3)\in \[0,4.8\]\times \[0,2.4\]\times\[0,0.6\]$mm$^3$ are considered.
The dimensions of the two-dimensional domain are set in such a way that the size and number of the finite elements used for the numerical simulations presented in section \ref{sec:numerical_results} lead, for the considered grain sizes, to good convergence for a moderate computational time.
By extension, the lengths of the three-dimensional solid are similar. 
Bimodal as well as monomodal distributions of the grain size based on the equivalent diameters $d=\{80,160,240\}\mu$m and the volume fractions of large grains $F_{LG}=\{0.25,0.50,0.75\}$ are considered.
This results in three monomodal and nine bimodal microstructures (\textit{i.e.} three combinations of equivalent diameters and three volume fractions of large grains) for both two-dimensional and three-dimensional cases.

\begin{figure}[!t]
  \centering
  \input{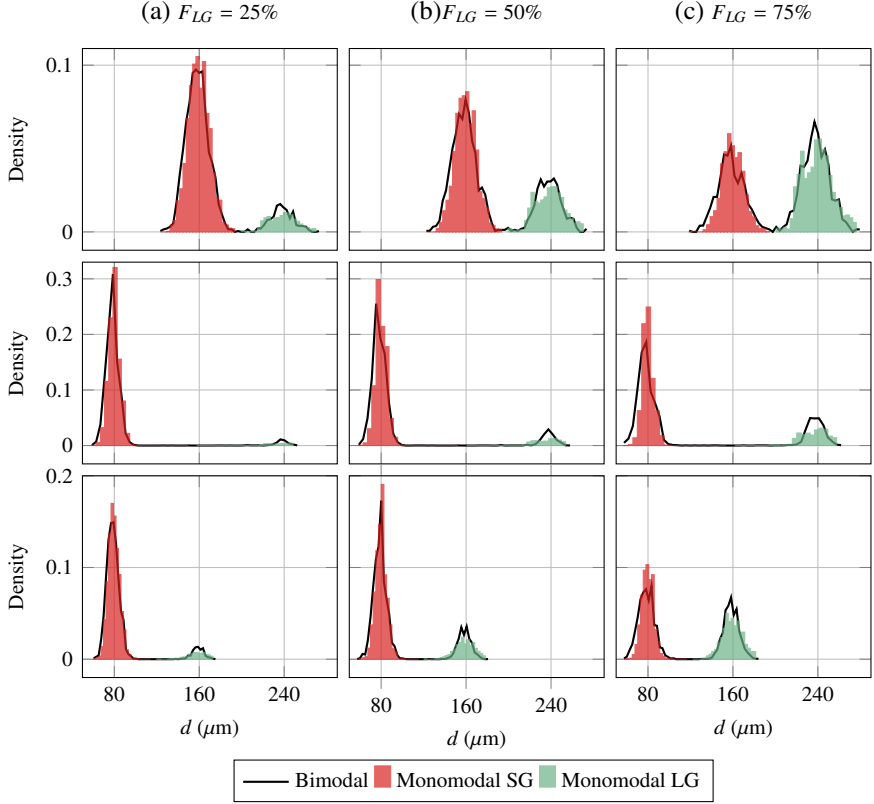}
  \caption{Comparison of the statistical distributions of equivalent diameter in two-dimensional bimodal microstructures with that of the corresponding monomodal microstructures for three volume fractions of large grains.
First row: $d^{SG}=160 \mu$m and $d^{LG}=240 \mu$m; Second row: $d^{SG}=80 \mu$m and $d^{LG}=240 \mu$m; Third row: $d^{SG}=80 \mu$m and $d^{LG}=160 \mu$m.}
  \label{fig:bimodal_distribution2D}
\end{figure}

\begin{figure}[!ht]
  \centering
  \input{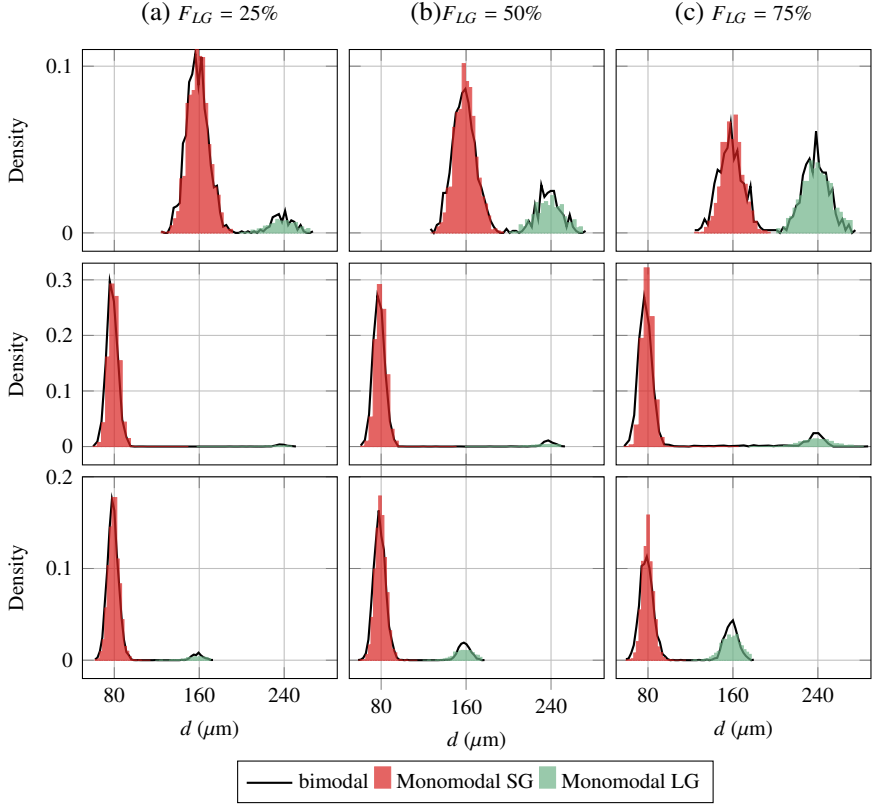}
  \caption{Comparison of the statistical distributions of equivalent diameter in three-dimensional bimodal microstructures with that of the corresponding monomodal microstructures for three volume fractions of large grains.
First row: $d^{SG}=160 \mu$m and $d^{LG}=240 \mu$m; Second row: $d^{SG}=80 \mu$m and $d^{LG}=240 \mu$m; Third row: $d^{SG}=80 \mu$m and $d^{LG}=160 \mu$m.}
  \label{fig:bimodal_distribution3D}
\end{figure}

The analysis of the equivalent diameter distributions in these twenty-four polycrystals 
\review{is presented} in Figures \ref{fig:bimodal_distribution2D} and \ref{fig:bimodal_distribution3D}, \review{which show} the comparison of the numerically constructed bimodal distributions \review{and} the associated monomodal ones for two and three space dimensions \review{respectively}.
The monomodal distributions depicted in the histograms are weighted with the numerical fraction of grains of the same family in the bimodal microstructure.
It is thus seen that the procedure described above for generating bimodal microstructures leads to a bimodal distribution of the equivalent diameter in which two modes can indeed clearly be identified. Moreover, even though bimodal and monomodal distributions do not overlap perfectly, both statistics show good agreement. \review{It finally enables validating the procedure for generating bimodal microstructures with \textsc{Neper}, which until then was not really explicit.}

\subsection{Validation of numerical autocorrelation functions} 
\label{sec:validation_W}

It is now proposed to check the conformity of the considered microstructures with regard to the additive partition of the spatial correlation function \eqref{eq:W_bimodal}.

At first glance, determining the two-point correlation function $W(r)$ for some polycrystal from equation \eqref{eq:W_mono} is rather complex.
It is however possible to approximate this function for numerical tessellations or EBSD data by following \cite{Man_Wfunction}, which is briefly recalled hereinafter.

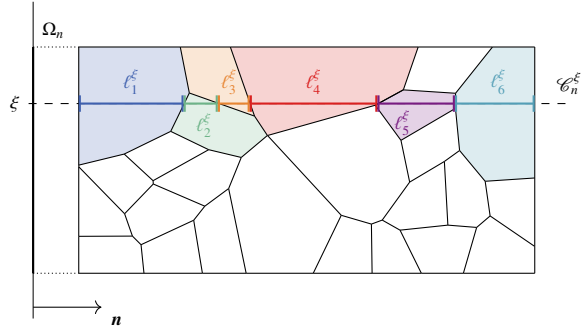
\begin{figure}[ht]
  \centering
  \begin{tikzpicture}[scale=3]
  \fill[Blue!20] (-0.0,0.477897832) -- (0.199457999,0.526834556) -- (0.40337674999999995,0.627121673) -- (0.485025716,0.7966810209999999) -- (0.445307982,1.0) -- (0.0,1.0) -- (-0.0,0.477897832) ;

  \fill[Green!20] (0.40337674999999995,0.627121673) -- (0.5698363869999999,0.546145668) -- (0.712781738,0.5118062179999999) -- (0.827491905,0.6089493399999999) --  (0.771045717,0.694587891) -- (0.485025716,0.7966810209999999) -- (0.40337674999999995,0.627121673);
  
  \fill[Orange!20] (0.771045717,0.694587891) -- (0.6649066530000001,1.0) -- (0.445307982,1.0) -- (0.485025716,0.7966810209999999) -- (0.771045717,0.694587891);
  
  \fill[Red!20] (1.488760156,1.0) -- (0.6649066530000001,1.0) -- (0.771045717,0.694587891) -- (0.827491905,0.6089493399999999) -- (0.8280318299999999,0.608841393) -- (1.3004973770000001,0.74264782) --  (1.423637324,0.810553195) -- (1.488760156,1.0);

  \fill[Purple!20] (1.423637324,0.810553195) -- (1.3004973770000001,0.74264782) -- (1.413218374,0.587482344) -- (1.652295606,0.7287195999999999) --  (1.6526799909999998,0.795509967) -- (1.423637324,0.810553195) ;

  \fill[Duck!20] (2.0,0.42112499299999995) -- (2.0,1.0) -- (1.8145845170000001,1.0) --  (1.6526799909999998,0.795509967) -- (1.652295606,0.7287195999999999) --  (1.670347311,0.542495372) -- (1.782558644,0.406574861) -- (2.0,0.42112499299999995) ;
  \draw (-0.0,0.477897832) -- (0.199457999,0.526834556); 
  \draw (0.199457999,0.526834556) -- (0.40337674999999995,0.627121673); 
  \draw (0.485025716,0.7966810209999999) -- (0.40337674999999995,0.627121673); 
  \draw (0.485025716,0.7966810209999999) -- (0.445307982,1.0); 
  \draw (0.445307982,1.0) -- (0.0,1.0); 
  \draw (0.0,1.0) -- (-0.0,0.477897832); 
  \draw (0.40337674999999995,0.627121673) -- (0.5698363869999999,0.546145668); 
  \draw (0.5698363869999999,0.546145668) -- (0.712781738,0.5118062179999999); 
  \draw (0.712781738,0.5118062179999999) -- (0.827491905,0.6089493399999999); 
  \draw (0.827491905,0.6089493399999999) -- (0.771045717,0.694587891); 
  \draw (0.771045717,0.694587891) -- (0.485025716,0.7966810209999999); 
\draw (0.771045717,0.694587891) -- (0.6649066530000001,1.0); 
  \draw (0.6649066530000001,1.0) -- (0.445307982,1.0); 
  \draw (1.488760156,1.0) -- (0.6649066530000001,1.0); 
  \draw (0.8280318299999999,0.608841393) -- (0.827491905,0.6089493399999999); 
  \draw (1.3004973770000001,0.74264782) -- (0.8280318299999999,0.608841393); 
  \draw (1.423637324,0.810553195) -- (1.3004973770000001,0.74264782); 
  \draw (1.488760156,1.0) -- (1.423637324,0.810553195); 
  \draw (1.423637324,0.810553195) -- (1.3004973770000001,0.74264782); 
  \draw (1.3004973770000001,0.74264782) -- (1.413218374,0.587482344); 
  \draw (1.413218374,0.587482344) -- (1.652295606,0.7287195999999999); 
  \draw (1.652295606,0.7287195999999999) -- (1.6526799909999998,0.795509967); 
  \draw (1.6526799909999998,0.795509967) -- (1.423637324,0.810553195); 
  \draw (2.0,0.42112499299999995) -- (2.0,1.0); 
  \draw (2.0,1.0) -- (1.8145845170000001,1.0); 
  \draw (1.6526799909999998,0.795509967) -- (1.8145845170000001,1.0); 
  \draw (1.670347311,0.542495372) -- (1.652295606,0.7287195999999999); 
  \draw (1.782558644,0.406574861) -- (1.670347311,0.542495372); 
  \draw (2.0,0.42112499299999995) -- (1.782558644,0.406574861); 
  
\draw (1.0285044399999999,-0.0) -- (1.286616793,-0.0); 
\draw (1.0285044399999999,-0.0) -- (1.151408547,0.256185781); 
\draw (1.151408547,0.256185781) -- (1.283654645,0.217320984); 
\draw (1.283654645,0.217320984) -- (1.286616793,-0.0); 
\draw (0.20112076099999998,0.26306081) -- (0.32697937699999996,0.39450635900000003); 
\draw (0.32697937699999996,0.39450635900000003) -- (0.552907196,0.29648161300000003); 
\draw (0.552907196,0.29648161300000003) -- (0.504179919,0.234334839); 
\draw (0.504179919,0.234334839) -- (0.273217871,0.12619283); 
\draw (0.273217871,0.12619283) -- (0.235473837,0.164252524); 
\draw (0.235473837,0.164252524) -- (0.20112076099999998,0.26306081); 
\draw (1.413218374,0.587482344) -- (1.4513872639999998,0.402880634); 
\draw (1.4513872639999998,0.402880634) -- (1.331070143,0.23503446099999997); 
\draw (1.331070143,0.23503446099999997) -- (1.283654645,0.217320984); 
\draw (1.151408547,0.256185781) -- (0.8280318299999999,0.608841393); 
\draw (0.5698363869999999,0.546145668) -- (0.326014052,0.409153091); 
\draw (0.326014052,0.409153091) -- (0.199457999,0.526834556); 
\draw (1.60362598,0.494114142) -- (1.4513872639999998,0.402880634); 
\draw (1.670347311,0.542495372) -- (1.60362598,0.494114142); 
\draw (1.60362598,0.494114142) -- (1.6189492719999998,0.212738028); 
\draw (1.6189492719999998,0.212738028) -- (1.4091714149999999,0.190556804); 
\draw (1.4091714149999999,0.190556804) -- (1.331070143,0.23503446099999997); 
\draw (-0.0,0.477897832) -- (-0.0,0.470633314); 
\draw (0.326014052,0.409153091) -- (0.32697937699999996,0.39450635900000003); 
\draw (0.20112076099999998,0.26306081) -- (-0.0,0.470633314); 
\draw (0.286085226,-0.0) -- (0.542368349,-0.0); 
\draw (0.504179919,0.234334839) -- (0.542368349,-0.0); 
\draw (0.286085226,-0.0) -- (0.273217871,0.12619283); 
\draw (0.734219707,-0.0) -- (1.0285044399999999,-0.0); 
\draw (0.734219707,-0.0) -- (0.6574491090000001,0.374341184); 
\draw (0.6574491090000001,0.374341184) -- (0.712781738,0.5118062179999999); 
\draw (0.6574491090000001,0.374341184) -- (0.552907196,0.29648161300000003); 
\draw (-0.0,0.163254283) -- (0.0,0.0); 
\draw (0.0,0.0) -- (0.286085226,-0.0); 
\draw (-0.0,0.163254283) -- (0.235473837,0.164252524); 
\draw (0.542368349,-0.0) -- (0.734219707,-0.0); 
\draw (1.8145845170000001,1.0) -- (1.488760156,1.0); 
\draw (2.0,0.0) -- (2.0,0.118535071); 
\draw (1.686981817,-0.0) -- (2.0,0.0); 
\draw (1.686981817,-0.0) -- (1.745231113,0.20257404799999998); 
\draw (1.745231113,0.20257404799999998) -- (1.804282256,0.25335205499999996); 
\draw (1.804282256,0.25335205499999996) -- (2.0,0.118535071); 
\draw (1.745231113,0.20257404799999998) -- (1.6189492719999998,0.212738028); 
\draw (1.782558644,0.406574861) -- (1.804282256,0.25335205499999996); 
\draw (1.502660976,-0.0) -- (1.686981817,-0.0); 
\draw (1.502660976,-0.0) -- (1.4091714149999999,0.190556804); 
\draw (1.286616793,-0.0) -- (1.502660976,-0.0); 
\draw (2.0,0.118535071) -- (2.0,0.42112499299999995); 
\draw (-0.0,0.470633314) -- (-0.0,0.163254283); 
\draw[ultra thin] (-0.2,-0.2) -- (-0.2,1.2);
\draw[densely dotted] (-0.2,-0.) -- (0.,0.);
\draw[densely dotted] (-0.2,1.) -- (0.,1.);
\draw[thick] (-0.2,-0.) -- (-0.2,1.);
\draw[dashed] (-0.22,.75) node[left] {\scriptsize $\xi$} -- (2.15,0.75) node [above] {\scriptsize $\Cscr^\xi_n$};

\draw[thin] (-0.2,-0.) -- (-0.2,1.) node[above right] {\scriptsize $\Omega_n$};
\draw[->] (-0.2,-0.15) -- (0.1,-0.15) node[below right] {\scriptsize $\vect{n}$};

\draw[densely dotted] (-0.1,0.75);
\draw[Blue,thick,|-|] (0.,0.75) -- (0.46,0.75) ;
\node[Blue,above] at (0.23,0.75) {\scriptsize $\ell_1^\xi$};
\draw[Green,thick,|-|] (0.46,0.75) -- (0.61,0.75);
\node[Green,below] at (0.55,0.75) {\scriptsize $\ell_2^\xi$};
\draw[Orange,thick,|-|] (0.61,0.75) -- (0.75,0.75);
\node[Orange,above] at (0.66,0.75) {\scriptsize $\ell_3^\xi$};
\draw[Red,thick,|-|] (0.75,0.75) -- (1.31,0.75);
\node[Red,above] at (1.03,0.75) {\scriptsize $\ell_4^\xi$};
\draw[Purple,thick,|-|] (1.31,0.75) -- (1.65,0.75);
\node[Purple,below] at (1.42,0.76) {\scriptsize $\ell_5^\xi$};
\draw[Duck,thick,|-|] (1.65,0.75) -- (2.,0.75);
\node[Duck,above] at (1.84,0.75) {\scriptsize $\ell_6^\xi$};

\end{tikzpicture}

  \caption{Schematic representation of the chords measure in a two-dimensional microstructure.}
  \label{fig:chord_length}
\end{figure}

Given an arbitrary direction $\vect{n} \in \Rbb^{dim}$, one defines $\Omega_n$ as the projection of $\Omega$ on the hyperplane through the origin and orthogonal to $\vect{n}$.
Moreover, $\Cscr^\xi_n$ are lines parallel to $\vect{n}$ starting from any point $\vect{\xi} \in \Omega_n$ (see figure \ref{fig:chord_length}).
Notice that the lines $\Cscr^\xi_n$ have to intersect a non-zero number of grains denoted as $N^g(\xi)$.  
The \textit{chord length} $\ell^\xi_i$ is defined as the length of the segment of $\Cscr^\xi_n$ contained in the $i$th grain.
Then, discretizing the subset $\Omega_n$ into $N_{\xi}$ points allows the approximation of the spatial correlation function as:
\begin{equation}
  \label{eq:Wfunc_general}
  W(r) \approx \sum_{i=1}^{N_\xi} \sum_{j=1}^{N^g(\xi_i)} \frac{\left\langle \ell_j^{\xi_i} -r \right\rangle^+}{\bar{\ell}}
\end{equation}
where $\bar{\ell}=\frac{\sum_{i=1}^{N_\xi} \sum_{j=1}^{N^g(\xi_i)} \ell_j^{\xi^i}}{\sum_{i=1}^{N_\xi} N^g(\xi_i)}$ is the mean chord length in $\Omega$ and $\left\langle \bullet \right\rangle^+ = \frac{\abs{\bullet} + \bullet }{2}$ is the positive part operator.

\begin{figure*}[ht]
  \centering
  \input{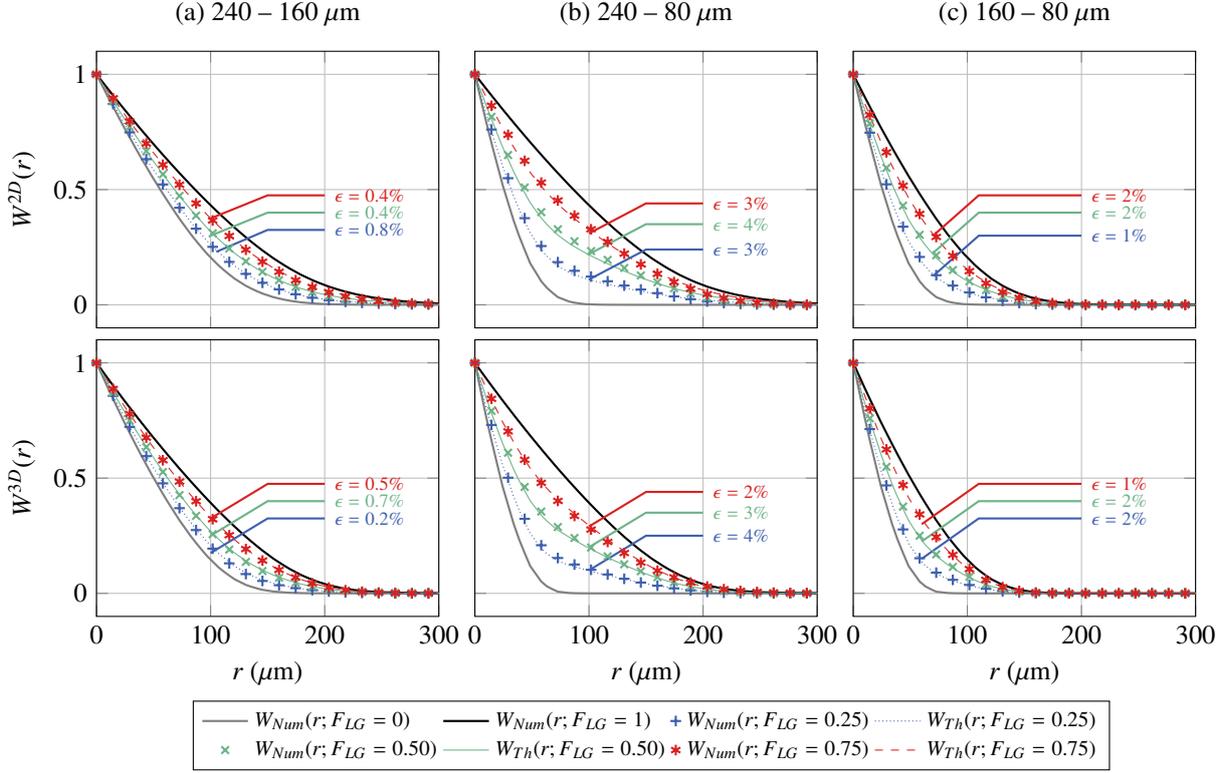}
\caption{\review{Comparison, for several volume fractions of large grains, between the spatial autocorrelation function for bimodal distributions microstructures and the theoretically expected weighted addition from two monomodal microstructures, with the L2 norm of the relative error between numerical and expected results $\epsilon$ annotated.} First row: two-dimensional cases; Second row: three-dimensional cases}
  \label{fig:W_bimodal}
\end{figure*}

Equation \eqref{eq:Wfunc_general} is used for the twenty-four numerical microstructures by superimposing lines parallel to the direction $\vect{e}_1$ to the crystal images.
The choice of the propagation direction $\vect{e}_1$ is made in accordance with the simulations of Section \ref{sec:numerical_results}.
It is worth noticing that the computation of the chord lengths is straightforward owing to the ability of \textsc{Neper} to rasterize tessellations\footnote{The pixels or voxels size is set to $h=1.2\times 10^{-5}$m, which leads to $N_\xi = 400$ in 2D and $N_\xi=10000$ in 3D.}.
Discrete values of \review{the} spatial correlation function are thus computed and $W(r)$ is at last reconstructed by means of a cubic spline function. \review{However, in the following (Figure \ref{fig:W_bimodal}), some $W(r)$ curves are plotted using discrete markers for the sake of comparison clarity.}

\review{On the other hand, the conformity of the considered bimodal microstructures in terms of  the spatial correlation function $W(r)$ with regard to theoretically expected  weighted addition from two monomodal microstructures given by \eqref{eq:W_bimodal} is looked at.
The autocorrelation functions reconstructed from the numerical grain data and denoted by $W_{Num}(r;F_{LG})$ are then} compared to the theoretically expected one $W_{Th}(r;F_{LG})$, which is defined as follows:
\begin{equation}
  \label{eq:th_expectedW}
   W_{Th}(r;F_{LG}) = F_{LG} W_{Num}(r;1) + (1-F_{LG}) W_{Num}(r;0)
\end{equation}
In (\ref{eq:th_expectedW}), the functions $W_{Num}(r;0)$ and $W_{Num}(r;1)$ correspond to the monomodal cases $F_{LG}=0$ and $F_{LG}=1$ and are reconstructed from the numerical grain data.
\review{Therefore, $W_{Th}(r;F_{LG})$ is in fact a theorectical-numerical hybrid function since it comes from a theoretical formula \eqref{eq:W_bimodal} but used with two numerically evaluated functions $W_{Num}(r;1)$ and $W_{Num}(r;0)$.} 

Figure \ref{fig:W_bimodal} shows \review{such comparisons} for the considered \review{bimodal} microstructures in two and three space dimensions.
In each plot of the figure, three volume fractions of large grains are considered for one combination of modal equivalent diameters and compared to the associated monomodal curves\review{, which} give rise to the bounding curves in each plot.

It can first be seen that the bimodal spatial correlation functions lie between the monomodal limits for each combination of modal diameters.
Note also that the lower (\textit{resp.} the higher) the volume fraction of large grains, the closer the curve is to the extreme case for which the sample is made of small (\textit{resp.} large) grains only.
This is indeed seen in figure \ref{fig:W_bimodal} since the bimodal curves get closer of the lower bound curve with decreasing $F_{LG}$ in any case.

Second, each bimodal curve is also annotated with the $L^2$ norm of the relative error between numerical and expected results for bimodal distributions, which is calculated as follows:
\begin{equation}
  \label{eq:error}
  \epsilon = \(\frac{\sum_i \[W_{Th}(r_i;F_{LG}) - W_{Num}(r_i;F_{LG})\]^2}{\sum_i W^2_{Th}(r_i;F_{LG})}\)^{1/2}
\end{equation}
Thus, one sees that the reconstructed autocorrelation functions are very close to the expected ones.
Notice however that the highest errors occur for the highest ratio between large and small modal diameters for both two and three-dimensional results, although they remain lower than $5\%$.

\begin{remark}
  The spline interpolation of the grain data presented here makes straightforward the computation of the attenuation coefficient \eqref{eq:alpha_integral} accounting for the morphology of the sample through equation \eqref{eq:generic_D}.
  It must however be emphasized that one has to ensure the fulfillment of the following condition \cite{Man_Wfunction}: 
  \begin{equation}
    \label{eq:Man_W'}
    W'(r=0) = -  \frac{\sum_{i=1}^{N_\xi}N^g(\xi_i)}{\bar{\ell} }
  \end{equation}
  which is a mathematical property of the autocorrelation function that may be not automatically satisfied by the numerical approximation.
  Such computations of the attenuation coefficient are proposed in section \ref{sec:decomposition_alpha}.
\end{remark}

\subsection{Material parameters}
\label{sec:material-parameters}
In the remainder of the paper, untextured and single-phase polycrystals with equiaxed grains of cubic symmetry are considered. 
Such polycrystals are modeled by setting to the grains a random orientation of the cubic axes with respect to the Cartesian basis, which can be done by choosing Euler angles ($\varphi_1,\phi,\varphi_2$) as:
\begin{equation}
  \label{eq:euler_angles}
  \varphi_1 = \text{random}\left[ 0,2\pi\right[, \quad \phi=\arccos(\text{random}\left[ -1,1\right]),\quad \varphi_2 = \text{random}\left[ 0,2\pi\right[
\end{equation}

On the other hand, the values of material parameters for both the homogeneous reference medium and the polycrystals are gathered in table \ref{tab:material}.
\begin{table}[ht]
  \centering
  \begin{tabular}{c|cccc}
  & $C_{1111} $(GPa) & $C_{1122} $(GPa) & $C_{1212} $(GPa)& $\rho $(kg/m$^3$) \\
  \hline
  Single crystallite in cubic axes & 134 & 110&36 &4428\\
  Equivalent homogeneous material &153 &100 &26.5 & 4428
\end{tabular}

  \caption{Material parameters for a BCC $\beta$-titanium metal \cite{MatParam}.}
  \label{tab:material}
\end{table}
The anisotropy degree for longitudinal and transverse waves of the considered polycrystalline material are low compared to unity \cite{Xue_2Dalpha}.
Therefore, the equations developed in \ref{sec:alpha_formula} for untextured, single-phase and weakly-scattering materials hold.
Furthermore, the elastic properties of the equivalent medium correspond to the Voigt-averaged material over the set of orientations, as required \cite{Stanke_Kino84}.


\section{Influence of bimodal distributions of the grain size on the attenuation}
\label{sec:procedure}
In this section, the unified theory of \textsc{Stanke} and \textsc{Kino} embedding sample-defined autocorrelation functions, as developed in section \ref{sec:theory}, is used for the computation of the analytical attenuation coefficient in bimodal microstructures.
This extension of the unified theory to distributions of the equivalent diameters that do not follow Poisson statistics allows the calculation of the attenuation coefficient for two as well as three-dimensional problems which, to the authors' knowledge, is not possible \review{in the 2D case} when employing \textsc{Weaver}'s model as in \cite{Ryzy2018,Pamel2018,Sha2018}.
Even though problems in two space dimensions may seem irrelevant for modeling experimental non-destructive testing, which are in essence three-dimensional, they are considered here so that the numerical simulations (as presented in section \ref{sec:numerical_results}) can be performed at an admissible computational cost.

The observations made on the analytical results in two and three space dimensions then motivate the development of a procedure aiming at characterizing bimodal distributions of the equivalent diameter from attenuation data, which is presented and illustrated in section \ref{sec:procedure_presentation}.

\subsection{Validation of the additive partition of sample-defined attenuation coefficients}
\label{sec:decomposition_alpha}

\begin{figure*}[!ht]
  \centering
  \begin{tikzpicture}
  \begin{groupplot}[group style={group size=3 by 2,horizontal sep=3ex,vertical sep=1ex,yticklabels at=edge left,ylabels at=edge left,xticklabels at=edge bottom,xlabels at=edge bottom}
    ,width=0.37\textwidth
    ,xlabel=$f$(MHz)
    ,ymin=0,xmin=0
    ,xmax=20
    ]
    
    \nextgroupplot[title={(a) 240 -- 160$\mu$m},ylabel=$\alpha^{2D}$(1/m),ymax=140]
    
    \addplot[black!50,thick] table[x expr=\thisrow{f}/1.e6,y=160mm] {pgfFigures/pgfFiles/bimodal_analytical2D.pgf};
    \addplot[black,thick] table[x expr=\thisrow{f}/1.e6,y=240mm] {pgfFigures/pgfFiles/bimodal_analytical2D.pgf};

    \addplot[Blue,only marks,mark repeat=3,mark=+,thick] table[x expr=\thisrow{f}/1.e6,y=240_160mm_25] {pgfFigures/pgfFiles/bimodal_analytical2D.pgf};
    \addplot[Blue,thin,densely dotted] table[x expr=\thisrow{f}/1.e6,y expr=0.75*\thisrow{160mm}+0.25*\thisrow{240mm}] {pgfFigures/pgfFiles/bimodal_analytical2D.pgf};

    \addplot[Green,only marks,mark repeat=3,mark=x,thick] table[x expr=\thisrow{f}/1.e6,y=240_160mm_50] {pgfFigures/pgfFiles/bimodal_analytical2D.pgf};
    \addplot[Green,thin] table[x expr=\thisrow{f}/1.e6,y expr=0.5*\thisrow{160mm}+0.5*\thisrow{240mm}] {pgfFigures/pgfFiles/bimodal_analytical2D.pgf};

    \addplot[Red,only marks,mark repeat=3,mark=asterisk,thick] table[x expr=\thisrow{f}/1.e6,y=240_160mm_75] {pgfFigures/pgfFiles/bimodal_analytical2D.pgf};
    \addplot[Red,thin, dashed] table[x expr=\thisrow{f}/1.e6,y expr=0.25*\thisrow{160mm}+0.75*\thisrow{240mm}] {pgfFigures/pgfFiles/bimodal_analytical2D.pgf};
    
    \node[right,Red] at (15,62.72) {\scriptsize $\epsilon=0.8\%$};
    \node[right,Green] at (15,71.08) {\scriptsize $\epsilon=0.4\%$};
    \node[right,Blue] at (15,79.62) {\scriptsize $\epsilon=0.3\%$};

    \nextgroupplot[title={(b) 240 -- 80 $\mu$m},ymax=140]

    \addplot[black!50,thick] table[x expr=\thisrow{f}/1.e6,y=80mm] {pgfFigures/pgfFiles/bimodal_analytical2D.pgf};
    \addplot[black,thick] table[x expr=\thisrow{f}/1.e6,y=240mm] {pgfFigures/pgfFiles/bimodal_analytical2D.pgf};

    \addplot[Blue,only marks,mark repeat=3,mark=+,thick] table[x expr=\thisrow{f}/1.e6,y=240_80mm_25] {pgfFigures/pgfFiles/bimodal_analytical2D.pgf};
    \addplot[Blue,thin,densely dotted] table[x expr=\thisrow{f}/1.e6,y expr=0.75*\thisrow{80mm}+0.25*\thisrow{240mm}] {pgfFigures/pgfFiles/bimodal_analytical2D.pgf};

    \addplot[Green,only marks,mark repeat=3,mark=x,thick] table[x expr=\thisrow{f}/1.e6,y=240_80mm_50] {pgfFigures/pgfFiles/bimodal_analytical2D.pgf};
    \addplot[Green,thin] table[x expr=\thisrow{f}/1.e6,y expr=0.5*\thisrow{80mm}+0.5*\thisrow{240mm}] {pgfFigures/pgfFiles/bimodal_analytical2D.pgf};

    \addplot[Red,only marks,mark repeat=3,mark=asterisk,thick] table[x expr=\thisrow{f}/1.e6,y=240_80mm_75] {pgfFigures/pgfFiles/bimodal_analytical2D.pgf};
    \addplot[Red,thin, dashed] table[x expr=\thisrow{f}/1.e6,y expr=0.25*\thisrow{80mm}+0.75*\thisrow{240mm}] {pgfFigures/pgfFiles/bimodal_analytical2D.pgf};

    \node[right,Red] at (15,68) {\scriptsize $\epsilon=3\%$};
    \node[right,Green] at (15,84.2) {\scriptsize $\epsilon=4\%$};
    \node[right,Blue] at (15,103.52) {\scriptsize $\epsilon=2\%$};

    \nextgroupplot[title={(c) 160 -- 80 $\mu$m},ymax=140]
    \addplot[black!50,thick] table[x expr=\thisrow{f}/1.e6,y=80mm] {pgfFigures/pgfFiles/bimodal_analytical2D.pgf};
    \addplot[black,thick] table[x expr=\thisrow{f}/1.e6,y=160mm] {pgfFigures/pgfFiles/bimodal_analytical2D.pgf};

    \addplot[Blue,only marks,mark repeat=3,mark=+,thick] table[x expr=\thisrow{f}/1.e6,y=160_80mm_25] {pgfFigures/pgfFiles/bimodal_analytical2D.pgf};
    \addplot[Blue,thin,densely dotted] table[x expr=\thisrow{f}/1.e6,y expr=0.75*\thisrow{80mm}+0.25*\thisrow{160mm}] {pgfFigures/pgfFiles/bimodal_analytical2D.pgf};

    \addplot[Green,only marks,mark repeat=3,mark=x,thick] table[x expr=\thisrow{f}/1.e6,y=160_80mm_50] {pgfFigures/pgfFiles/bimodal_analytical2D.pgf};
    \addplot[Green,thin] table[x expr=\thisrow{f}/1.e6,y expr=0.5*\thisrow{80mm}+0.5*\thisrow{160mm}] {pgfFigures/pgfFiles/bimodal_analytical2D.pgf};

    \addplot[Red,only marks,mark repeat=3,mark=asterisk,thick] table[x expr=\thisrow{f}/1.e6,y=160_80mm_75] {pgfFigures/pgfFiles/bimodal_analytical2D.pgf};
    \addplot[Red,thin, dashed] table[x expr=\thisrow{f}/1.e6,y expr=0.25*\thisrow{80mm}+0.75*\thisrow{160mm}] {pgfFigures/pgfFiles/bimodal_analytical2D.pgf};

    \node[right,Red] at (15,97.37) {\scriptsize $\epsilon=0.9\%$};
    \node[right,Green] at (15,106.24) {\scriptsize $\epsilon=0.9\%$};
    \node[right,Blue] at (15,115.28) {\scriptsize $\epsilon=0.7\%$};

    \nextgroupplot[ylabel=$\alpha^{3D}$(1/m),ymax=225]
    \addplot[black!50,thick] table[x expr=\thisrow{f}/1.e6,y=160mm] {pgfFigures/pgfFiles/bimodal_analytical3D.pgf};
    \addplot[black,thick] table[x expr=\thisrow{f}/1.e6,y=240mm] {pgfFigures/pgfFiles/bimodal_analytical3D.pgf};

    \addplot[Blue,only marks,mark repeat=3,mark=+,thick] table[x expr=\thisrow{f}/1.e6,y=240_160mm_25] {pgfFigures/pgfFiles/bimodal_analytical3D.pgf};
    \addplot[Blue,thin,densely dotted] table[x expr=\thisrow{f}/1.e6,y expr=0.75*\thisrow{160mm}+0.25*\thisrow{240mm}] {pgfFigures/pgfFiles/bimodal_analytical3D.pgf};

    \addplot[Green,only marks,mark repeat=3,mark=x,thick] table[x expr=\thisrow{f}/1.e6,y=240_160mm_50] {pgfFigures/pgfFiles/bimodal_analytical3D.pgf};
    \addplot[Green,thin] table[x expr=\thisrow{f}/1.e6,y expr=0.5*\thisrow{160mm}+0.5*\thisrow{240mm}] {pgfFigures/pgfFiles/bimodal_analytical3D.pgf};

    \addplot[Red,only marks,mark repeat=3,mark=asterisk,thick] table[x expr=\thisrow{f}/1.e6,y=240_160mm_75] {pgfFigures/pgfFiles/bimodal_analytical3D.pgf};
    \addplot[Red,thin, dashed] table[x expr=\thisrow{f}/1.e6,y expr=0.25*\thisrow{160mm}+0.75*\thisrow{240mm}] {pgfFigures/pgfFiles/bimodal_analytical3D.pgf};

    \node[right,Red] at (15,128.82) {\scriptsize $\epsilon=0.6\%$};
    \node[right,Green] at (15,153.86) {\scriptsize $\epsilon=1\%$};
    \node[right,Blue] at (15,181.40) {\scriptsize $\epsilon=0.1\%$};
    
    \nextgroupplot[,ymax=225]
    \addplot[black!50,thick] table[x expr=\thisrow{f}/1.e6,y=80mm] {pgfFigures/pgfFiles/bimodal_analytical3D.pgf};
    \addplot[black,thick] table[x expr=\thisrow{f}/1.e6,y=240mm] {pgfFigures/pgfFiles/bimodal_analytical3D.pgf};

    \addplot[Blue,only marks,mark repeat=3,mark=+,thick] table[x expr=\thisrow{f}/1.e6,y=240_80mm_25] {pgfFigures/pgfFiles/bimodal_analytical3D.pgf};
    \addplot[Blue,thin,densely dotted] table[x expr=\thisrow{f}/1.e6,y expr=0.75*\thisrow{80mm}+0.25*\thisrow{240mm}] {pgfFigures/pgfFiles/bimodal_analytical3D.pgf};

    \addplot[Green,only marks,mark repeat=3,mark=x,thick] table[x expr=\thisrow{f}/1.e6,y=240_80mm_50] {pgfFigures/pgfFiles/bimodal_analytical3D.pgf};
    \addplot[Green,thin] table[x expr=\thisrow{f}/1.e6,y expr=0.5*\thisrow{80mm}+0.5*\thisrow{240mm}] {pgfFigures/pgfFiles/bimodal_analytical3D.pgf};

    \addplot[Red,only marks,mark repeat=3,mark=asterisk,thick] table[x expr=\thisrow{f}/1.e6,y=240_80mm_75] {pgfFigures/pgfFiles/bimodal_analytical3D.pgf};
    \addplot[Red,thin, dashed] table[x expr=\thisrow{f}/1.e6,y expr=0.25*\thisrow{80mm}+0.75*\thisrow{240mm}] {pgfFigures/pgfFiles/bimodal_analytical3D.pgf};

    \node[right,Red] at (15,128.5) {\scriptsize $\epsilon=0.5\%$};
    \node[right,Green] at (15,149) {\scriptsize $\epsilon=3\%$};
    \node[right,Blue] at (15,172.5) {\scriptsize $\epsilon=2\%$};

    \nextgroupplot[legend style={at={($(.65,-0.4)+(0cm,0.35cm)$)},legend columns=4},ymax=225]
    
    \addplot[black!50,thick] table[x expr=\thisrow{f}/1.e6,y=80mm] {pgfFigures/pgfFiles/bimodal_analytical3D.pgf};
    \addplot[black,thick] table[x expr=\thisrow{f}/1.e6,y=160mm] {pgfFigures/pgfFiles/bimodal_analytical3D.pgf};

    \addplot[Blue,only marks,mark repeat=3,mark=+,thick] table[x expr=\thisrow{f}/1.e6,y=160_80mm_25] {pgfFigures/pgfFiles/bimodal_analytical3D.pgf};
    \addplot[Blue,thin,densely dotted] table[x expr=\thisrow{f}/1.e6,y expr=0.75*\thisrow{80mm}+0.25*\thisrow{160mm}] {pgfFigures/pgfFiles/bimodal_analytical3D.pgf};

    \addplot[Green,only marks,mark repeat=3,mark=x,thick] table[x expr=\thisrow{f}/1.e6,y=160_80mm_50] {pgfFigures/pgfFiles/bimodal_analytical3D.pgf};
    \addplot[Green,thin] table[x expr=\thisrow{f}/1.e6,y expr=0.5*\thisrow{80mm}+0.5*\thisrow{160mm}] {pgfFigures/pgfFiles/bimodal_analytical3D.pgf};

    \addplot[Red,only marks,mark repeat=3,mark=asterisk,thick] table[x expr=\thisrow{f}/1.e6,y=160_80mm_75] {pgfFigures/pgfFiles/bimodal_analytical3D.pgf};
    \addplot[Red,thin, dashed] table[x expr=\thisrow{f}/1.e6,y expr=0.25*\thisrow{80mm}+0.75*\thisrow{160mm}] {pgfFigures/pgfFiles/bimodal_analytical3D.pgf};

    \node[right,Red] at (15,215.) {\scriptsize $\epsilon=2\%$};
    \node[right,Green] at (15,203.6) {\scriptsize $\epsilon=2\%$};
    \node[below right,Blue] at (15,201.4) {\scriptsize $\epsilon=2\%$};

    \addlegendentry{$\alpha_{Th}$ $F_{LG}=0$}
    \addlegendentry{$\alpha_{Th}$ $F_{LG}=1$}

    \addlegendentry{$\alpha_{Th}$ $F_{LG}=0.25$}
    \addlegendentry{Expected $F_{LG}=0.25$}

    \addlegendentry{$\alpha_{Th}$ $F_{LG}=0.50$}
    \addlegendentry{Expected $F_{LG}=0.50$}

    \addlegendentry{$\alpha_{Th}$ $F_{LG}=0.75$}
    \addlegendentry{Expected $F_{LG}=0.75$}

  \end{groupplot}
\end{tikzpicture}

  \caption{Analytical sample-based attenuation curves for \review{two- and} three-dimensional microstructures: comparison between bimodal and monomodal distribution of the grain size and the expected results based on equation \eqref{eq:bimodal_decomposition}.}
  \label{fig:alphaTH_bimodal}
\end{figure*}
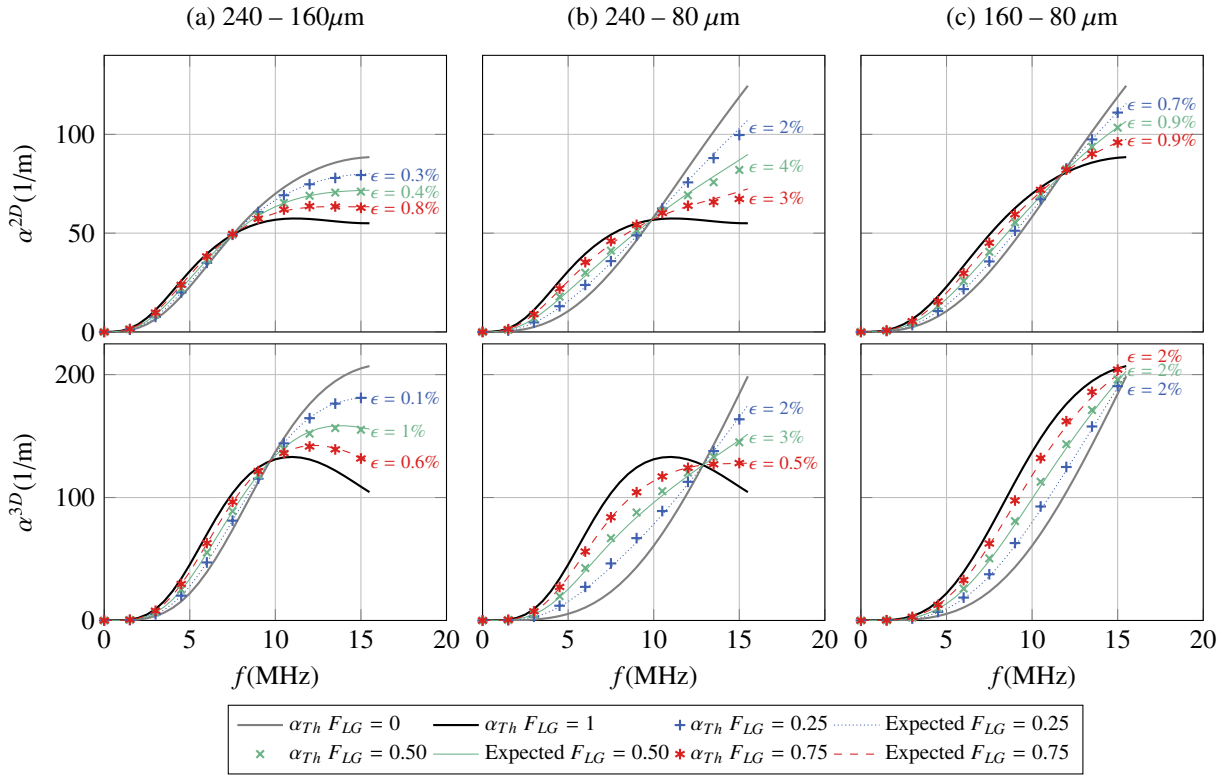

The formulas presented in \ref{sec:alpha_formula} allow the computation of analytical attenuation curves that take into account the autocorrelation functions determined as presented in the previous section.
Figure \ref{fig:alphaTH_bimodal} shows the longitudinal frequency-dependent attenuation coefficient for the two and three-dimensional bimodal microstructures considered so far.
Once again, color markers are used for bimodal microstructures while thick solid grey or black lines correspond to the associated monomodal ones. 
Moreover, the expected results coming from the additive partition of the attenuation coefficient \eqref{eq:bimodal_decomposition} are depicted using thin color lines for comparison purposes, the $L^2$ norm of the relative errors being also reported.

As for the spatial correlation function, one sees that the higher the volume fraction of large grains, the closer the curve is to the extreme case for which the sample is made of large grains only (and conversely).
Next, the error computed between the theoretical evaluation of the attenuation coefficients for bimodal microstructures and the convex combination \eqref{eq:bimodal_decomposition} is always lower than or equal to $6\%$, which shows good agreement.
As a result, in each figure all the curves cross at the same point as expected with the convex combination \eqref{eq:bimodal_decomposition}.
%

In light of the above results, it seems possible to estimate the volume fraction of large grains in a bimodal microstructure based on the attenuation curve.
Let us assume that the frequency-dependent attenuation coefficient corresponding to the modal equivalent diameters are known \textit{a priori}, then, the volume fraction of large grains is in fact the solution of equation \eqref{eq:bimodal_decomposition}. 
Nevertheless, such an approach is based on the knowledge of the modal equivalent diameters and the corresponding attenuation curves, the former being the object of non-destructive testing as well.
The object of the following is to improve the outlined approach in order to deduce the volume fraction of large grains as well as the two modal equivalent diameters from the attenuation data of a bimodal microstructure.

\subsection{Inverse analysis: characterization of bimodal distributions of the grain size}
\label{sec:procedure_presentation}
We denote attenuation data collected experimentally on a bimodal distribution as the vector $\vect{\mathrm{A}}$ whose $N_f$ lines correspond to frequency values.
The results of the previous section confirm that these data can be written as:
\begin{equation}
  \label{eq:bimodal_approximation}
  \mathrm{A}_i  = F_{LG}\alpha(f_i,d^{LG}) + (1-F_{LG})\alpha(f_i,d^{SG})
\end{equation}
in which $f_i$ is the $i$-th frequency value, and $\alpha(f_i,\bullet)$ is well representative of monomodal microstructures.
Then, the set of parameters $\{F_{LG},d^{SG},d^{LG}\}$ that minimizes the errors between the left and right-hand sides of equation \eqref{eq:bimodal_approximation} can be determined by using some well-known optimization procedures, such as nonlinear least squares.
The problem therefore reads:
\begin{align}
  \label{eq:optimization_problem}
  &\{F_{LG},d^{SG},d^{LG}\} = \arg \underset{F,d_1,d_2}{\min} \left\lbrace \sum_{i=1}^{N_f} \[A_i - g^{B}(f_i,F,d_1,d_2) \]^2 \right\rbrace \\
  & g^{B}(f_i,F,d_1,d_2) =F\alpha(f_i,d_2)  + (1-F)\alpha(f_i,d_1)
\end{align}
This however let some freedom in the choice of $\alpha(f,d)$ in the model function $g^{B}$.

Employing a minimization procedure raises the question of finding local or global minima and hence, ensuring the solution's uniqueness.
Nevertheless, it is believed that the initial guess of the parameters can be controlled so as to prevent any problem.
Indeed, since the monomodal curves and the bimodal one must cross at the same point, initial equivalent diameters $d_0^{SG}$ and $d_0^{LG}$ can be deduced from the solution of the following problem:
\begin{align}
  \label{eq:optimization_problem}
  &d_0^{SG},d_0^{LG} = \arg \underset{d_1,d_2}{\min} \left\lbrace \abs{\alpha(\widetilde{f},d_1)-  \widetilde{A}} + \abs{\alpha(\widetilde{f},d_2) - \widetilde{A}}   \right\rbrace \\
  & \widetilde{f} \in \{f : \alpha(f,d_1)=\alpha(f,d_2) \} 
\end{align}
in which $\widetilde{A}$ is the interpolation of $\vect{A}$ at frequency $\widetilde{f}$.
The use of such an initial guess in the least squares approach should restrict the parameter space so that the solution of optimization \eqref{eq:optimization_problem} is satisfactory. 

\review{
  It is worth noticing this crossing point indicates that the same attenuation coefficient is obtained for two different grain sizes.
  Therefore, regarding the frequency at which such a crossing point is observed for two monomodal microstructures $SG$ and $LG$, we believe that it should not correspond to the same scattering domain for the two microstructures.
  Indeed, according to the form  $\alpha \propto d^{n-1}f^n$, it is seen that the $LG$ curve is above the $SG$ one for a given scattering domain.
  The intersection can then occur if the former presents an inflection, which would be due to a change in its scattering domain.
  This for example means that if the frequency of the crossing point is in the Rayleigh domain for the $SG$ microstructure, it should at least be in the Rayleigh-to-stochsatic transition zone for the $LG$ microstructure.
  Furthermore, Figure \ref{fig:alphaTH_bimodal} shows that the crossing occurs at a higher frequency in the 3D case than in the 2D case, which is coherent with the conclusion of our previous work stating that the transition starts at higher frequencies in the 3D case than in the 2D case \cite{Xue_2Dalpha}.}

\begin{remark}
  An equivalent monomodal microstructure can also be found by setting $F=1$ in the model function $g^{B}(f,d)$.
  The problem then becomes:
  \begin{equation}
    \label{eq:optimization_problem_mono}
    \bar{d} = \arg \underset{d}{\min}  \left\lbrace \sum_{i=1}^{N_f} \[A_i - \alpha(f_i,d) \]^2 \right\rbrace 
  \end{equation}
  In what follows, optimizations \eqref{eq:optimization_problem} and \eqref{eq:optimization_problem_mono} are referred to as \textit{bimodal fitting} and \textit{monomodal fitting} respectively.  
\end{remark}

\begin{remark}
  Although this work focuses on bimodal distributions, the presented approach could be generalized to the characterization of ``spread'' monomodal distributions of the equivalent diameter. 
  Indeed, given the form of $\alpha$ in equation \eqref{eq:multiphase_decomposition}, it should be possible to approximate a grain size distribution by seeking the equivalent diameters and volume fractions of $N$ families of grains.
  However, such an approach, which could allow to generalize the inversion proposed in \cite{nicoletti1997}, is the object of future works.
\end{remark}

As a first illustration, we propose to characterize two and three-dimensional bimodal microstructures with the same morphology as before, for which the analytical attenuation curves are viewed as experimental data. 
%
%

In order to construct the model function $g^B$ upon a representative function $\alpha(f,d)$, we propose to use theoretical sample-based evaluations of the attenuation curve in monomodal microstructures having similar grain shapes as the bimodal ones. 
Since such curves involve sample-defined spatial autocorrelation, they can only be known for discrete equivalent diameters, in contrast with those resulting from the exponential form.
Nevertheless, this limitation can be overcome by reconstructing a surface by means of bivariate spline interpolation \cite{BivariateSpline}.
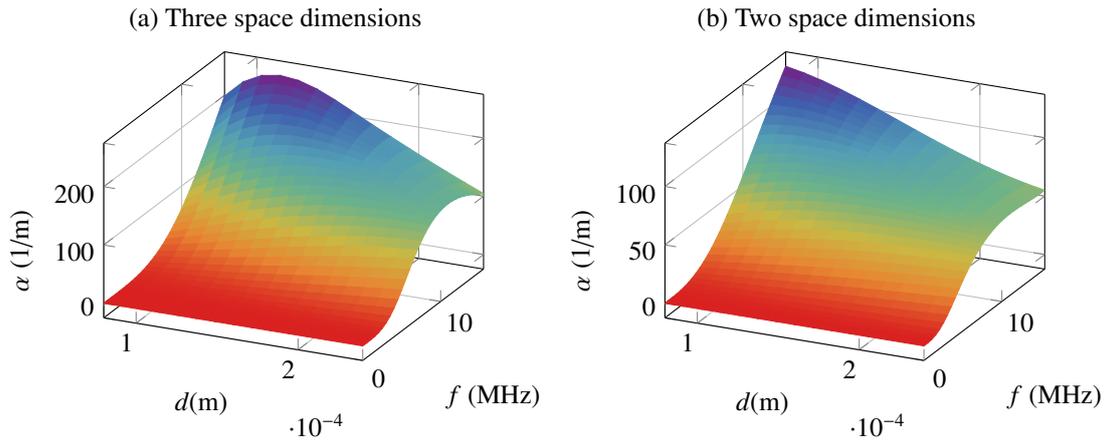
\begin{figure}[!ht]
  \centering
\subcaptionbox{Three space dimensions}{\begin{tikzpicture}
  \begin{axis}[xlabel=$d $(m),ylabel=$f$ (MHz),zlabel=$\alpha$ (1/m)
    ,width=0.40\linewidth
    ]
    \addplot3[surf,shader=flat] table[x=d,y expr=\thisrow{f}/1.e6,z=alpha] {pgfFigures/pgfFiles/alpha_analyticalSurface3D.pgf};
  \end{axis}
\end{tikzpicture}
}
\subcaptionbox{Two space dimensions}{\begin{tikzpicture}
  \begin{axis}[xlabel=$d $(m),ylabel=$f$ (MHz),zlabel=$\alpha$ (1/m)
    ,width=0.4\linewidth
    ]
    \addplot3[surf,shader=flat] table[x=d,y expr=\thisrow{f}/1.e6,z=alpha] {pgfFigures/pgfFiles/alpha_analyticalSurface2D.pgf};
  \end{axis}
\end{tikzpicture}
}
  \caption{Sample-based analytical attenuation coefficient in the ($d$,$f$,$\alpha$) space for monomodal microstructures.}
  \label{fig:analyticalBimodal}
\end{figure}
Figure \ref{fig:analyticalBimodal} shows reconstructions based on the attenuation results of the two and three-dimensional microstructures of equivalent diameters $d=\{80,120,160,200,240\}\mu$m using cubic interpolation in both directions $d$ and $f$.
The set of bimodal microstructures considered is then composed of those already considered plus one for which the associated monomodal results are not known, that is $F_{LG}=0.35$, $d^{LG}=190 \mu m$, $d^{SG}=90 \mu m$. 

First, each microstructure is assumed to be monomodal and problem \eqref{eq:optimization_problem_mono} is solved.
The results of that optimization are shown in tables \ref{tab:diameqTH_bimod2D} and \ref{tab:diameqTH_bimod3D} for two-dimensional and three-dimensional cases.
For each polycrystal, the $L^2$ norm of the relative error between the data and the optimized results, defined as: 
\begin{equation}
  \label{eq:error_mono}
  \epsilon_{Mono} = \(\frac{\sum_{i=1}^{N_f} \(A_i - \alpha(f_i,\bar{d})\)^2}{\sum_{k=1}^{N_f} A_k^2}\)^{1/2}
\end{equation}
is also reported.

Next, bimodal fittings \eqref{eq:optimization_problem}, whose optimized parameters are also gathered in the tables, are performed.
In these cases, the error is computed as:
\begin{equation}
  \label{eq:error_bimodal}
  \epsilon_{Bi} = \(\frac{\sum_{i=1}^{N_f} \(A_i - g^B(f_i,F_{LG},d^{SG},d^{LG})\)^2}{\sum_{k=1}^{N_f} A_k^2}\)^{1/2}
\end{equation}

In most cases, the optimal equivalent diameter resulting from monomodal fittings is close to the weighted average: $F_{LG}d^{LG} + (1-F_{LG})d^{SG}$.
\begin{table}[ht]
  \centering
  
\begin{tabular}{ccc|cc|cccc}
  \multicolumn{3}{c|}{Input parameters} & \multicolumn{2}{c|}{Monomodal fitting} & \multicolumn{4}{c}{Bimodal fitting}\\
  $F_{LG}$ (\%)& $d^{SG}$ & $d^{LG}$ & $d_{opt}$ & $\epsilon_{Mono}$ (\%) &  $F^{opt}_{LG}$ (\%)& $d_{opt}^{SG}$ & $d^{LG}_{opt}$& $\epsilon_{Bi}$ (\%) \\
  \hline
  25 & 160 & 240 & 179 & 2.0 & 36 & 155 & 225 & $5.8\times 10^{-2}$\\
  50 & 160 & 240 & 199 & 2.1 & 60 & 154 & 231 & $3.2\times 10^{-2}$\\
  75 & 160 & 240 & 218 & 1.2 & 90 & 138 & 228 & $1.0\times 10^{-1}$\\
  \hline
  25 & 80 & 240 & 136 & 14 & 30 & 80 & 237 & $2.9\times 10^{-1}$\\
  50 & 80 & 240 & 184 & 12 & 58 & 80 & 235 & $5.1\times 10^{-1}$\\
  75 & 80 & 240 & 214 & 6.1 & 83 & 80 & 233 & $5.3\times 10^{-1}$\\

  \hline
  25 & 80 & 160 & 94 & 4.7 & 29 & 80 & 153 & $1.3\times 10^{-1}$\\
  50 & 80 & 160 & 127 & 5.5  & 56 & 80 & 155 & $1.3\times 10^{-1}$\\
  75 & 80 & 160 & 145 & 2.8  & 81 & 80 & 155 & $6.8\times 10^{-2}$\\
  \hline
  35 & 90 & 190 & 134 & 6.6 & 38 & 91 & 191 & $2.2\times 10^{-1}$
\end{tabular}

  \caption{Results of monomodal and bimodal fittings for two-dimensional bimodal microstructures.}
  \label{tab:diameqTH_bimod2D}
\end{table}
\begin{table}[ht]
  \centering
  
\begin{tabular}{ccc|cc|cccc}
  \multicolumn{3}{c|}{Input parameters} & \multicolumn{2}{c|}{Monomodal fitting} & \multicolumn{4}{c}{Bimodal fitting}\\
  $F_{LG}$ (\%)& $d^{SG}$ & $d^{LG}$ & $d_{opt}$ & $\epsilon_{Mono}$ (\%) &  $F^{opt}_{LG}$ (\%)& $d_{opt}^{SG}$ & $d^{LG}_{opt}$& $\epsilon_{Bi}$ (\%) \\
  \hline
  25 & 160 & 240 & 180 & 3.2 & 27 & 159 & 238 & $3.9\times 10^{-2}$\\
  50 & 160 & 240 & 203 & 4.5 & 54 & 156 & 240 & $1.9\times 10^{-2}$\\
  75 & 160 & 240 & 223 & 3.0 & 77 & 157 & 240 & $2.7\times 10^{-2}$\\
  \hline
  25 & 80 & 240 & 209 & 42 & 29 & 80 & 237 & $8.2\times 10^{-2}$\\
  50 & 80 & 240 & 224 & 26 & 55 & 80 & 237 & $1.1\times 10^{-1}$\\
  75 & 80 & 240 & 234 & 14 & 75 & 80 & 240 & $4.2\times 10^{-2}$\\

  \hline
  25 & 80 & 160 & 87 & 8.4 & 29 & 80 & 159 & $2.7\times 10^{-2}$\\
  50 & 80 & 160 & 97& 14 & 54 & 80 & 159 & $6.6\times 10^{-2}$\\
  75 & 80 & 160 & 154 & 8.0 & 77 & 84 & 159 & $4.5\times 10^{-2}$\\
  \hline
  35 & 90 & 190 & 154 & 22 & 42 & 87  & 183 & $2.9\times 10^{-2}$
\end{tabular}

  \caption{Results of monomodal and bimodal fittings for three-dimensional bimodal microstructures.}
  \label{tab:diameqTH_bimod3D}
\end{table}
Moreover, the monomodal errors are in any cases far greater than that of bimodal fittings, the former being between $10$ and $10^2$ times larger than the latter.
Furthermore, note that the higher the ratio $d^{LG}/d^{SG}$, the higher the monomodal error.
Therefore, a large value of $\epsilon_{Mono}$ gives an indication about the non-monomodal nature of the actual distribution of the equivalent diameter.
On the other hand, the parameters resulting from the bimodal fitting are in general close to the input data, even though they are not identical to the input data.
These gaps may be reduced by using additional monomodal attenuation curves to reconstruct the surface $\alpha(f,d)$.
At last, analogously to the result of monomodal fittings, one sees that the highest bimodal errors occur for high values of the ratio $d^{LG}/d^{SG}$.

The approach proposed here for the characterization of bimodal distribution of the grain size provides very encouraging results.
On the one hand, using a set of attenuation curves in order to reconstruct an attenuation model $\alpha(f,d)$ is something that can be done experimentally or numerically.
This point allows to avoid gaps between collected data resulting from multiple-scattering and the analytical results accounting for single-scattering.
%
On the other hand, solving an optimization problem over the whole frequency domain enables to get rid of the problems related to the frequency exponent that varies depending on the scattering region.

Nevertheless, the above illustrations emphasize that the proposed bimodal optimization can be improved since the actual morphological parameters are not exactly recovered.
Additional effort should then be done on the model function and more specifically on the reconstruction of the surface $\alpha(f,d)$.
Indeed, a cubic interpolation is performed although it is well known that the dependency of the attenuation coefficient on $d$ changes depending on the scattering region. 
Two solutions could allow to circumvent this problem: (1) do a bivariate spline approximation with a polynomial degree in direction $d$ varying with respect to the scattering region; (2) use more discrete equivalent diameters for the computation of reference attenuation curves in monomodal microstructures.
Notice that the \review{same} remark holds for the frequency dependency\review{, which is solved in this work using option (2). Indeed}, numerous sample points are used in the frequency range so that the aforementioned approximation error is avoided.


\section{Numerical modeling of a two-dimensional problem}
\label{sec:numerical_results}
The procedure presented in the previous section is now applied to numerical attenuation data in two space dimensions that are computed in polycrystals whose material parameters have been presented in section \ref{sec:material-parameters} (see table \ref{tab:material}).
This constitutes a first step towards the application of the approach to experimental characterization of bimodal microstructures.
\subsection{Properties of the continuum problem}

\begin{figure}[ht]
  \centering
  \begin{tikzpicture}[scale=0.7]
  \draw[thick] (0,0) --(6,0)--(6,3)--(0,3)--(0,0);
  \foreach \x in {0.,0.25,...,3} 
  \draw[>=stealth,->] (-0.5,\x)--(0.,\x);
  
  \node(a)at(-2.5,1.5){ \scriptsize $\tens{\sigma}\cdot\vect{e}_1=\matrice{ \sigma_d(t) \\0\\0}$}; 
  \draw(0,3.4)--(6.,3.4);
  \draw(0,-.4)--(6.,-.4);
  
  \foreach \x in {0.5,1.,...,5.5} 
  \draw(\x,-0.2)circle(0.2);
  \foreach \x in {0.5,1.,...,5.5} 
  \draw(\x,3.2)circle(0.2);
  
  \fill [pattern=north east lines](0.,-0.4)rectangle+(6.,-0.4);
  \fill [pattern=north east lines](0.,3.4)rectangle+(6.,0.4);
  
  \draw[>=stealth,<->](5.9,0)--node[left=1pt]{\footnotesize $h= \:4.8 \times 10^{-3}$ m}(5.9,3);
  \draw[>=stealth,<->](0,0.2)--node[above=1pt]{\footnotesize  $l= \: 9.6 \times 10^{-3}$ m}(6,0.2);
  \draw[>=stealth,->](-2.5,-1.)--(-1.5,-1.)node(a)[anchor=north]{\footnotesize  $\vect{e}_1$};
  \draw[>=stealth,->](-2.5,-1.)--(-2.5,0)node(a)[anchor=south]{\footnotesize  $\vect{e}_2$};

\end{tikzpicture}
  \caption{Geometry and loading conditions of the considered problem}
  \label{fig:problem_setting}
\end{figure}
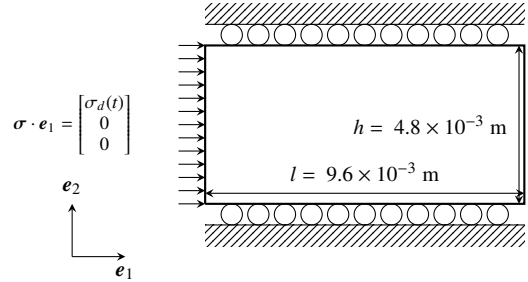

We consider a two-dimensional domain made of a polycrystalline material, submitted to a time-varying traction force on its left end as depicted in figure \ref{fig:problem_setting}.
This signal is taken as a sum of two Ricker signals varying with frequency content centered to $5$MHz and $10$MHz, which is representative of a \review{broad frequency range} pulse.
For such an external loading, the valid frequency domain is limited to a range $4$--$16$MHz, which leads to a range from $375\mu$m to $1.5$mm for wavelengths \review{for the $\beta$-titanium considered hereinafter} \cite{Xue_2Dalpha}.
Hence, the Rayleigh region and the Rayleigh-to-stochastic transition domain are covered with respect to the considered grain sizes.

As before, monomodal and bimodal distributions of the equivalent diameter, whose characteristics are shown in table \ref{tab:nGrains}, are considered.
\begin{table}[ht]
  \centering
  \begin{tabular}{l|ccc}
  $d$($\mu$m)   & $F_{LG}$ (\%) & $n^G$ & Number of samples  \\
  \hline
  80   & --- & 9167  & 5 \\  
  120   & --- & 4074  & 10 \\  
  160  & --- & 2291  & 10  \\  
  200  & --- & 1466  & 10  \\  
  240  & --- & 1018  & 20 \\
  \hline
  240--80 & \begin{tabular}{c}
               25 \\ 50 \\ 75 
               \end{tabular} & \begin{tabular}{c}
               7129 \\ 5092 \\ 3054  
               \end{tabular} & 10 \\
  \hline
  240--160 & \begin{tabular}{c}
               25 \\ 50 \\ 75 
               \end{tabular} & \begin{tabular}{c}
               1972 \\ 1654 \\ 1335  
               \end{tabular} & 10 \\
  \hline
  160--80 & \begin{tabular}{c}
               25 \\ 50 \\ 75 
               \end{tabular} & \begin{tabular}{c}
                7447 \\5728 \\ 4009   
               \end{tabular} & 10 \\
  \hline
  190--90  & 35 & 5276  & 10 \\
  \hline
\end{tabular}


  \caption{Number of grains in the considered microstructures and number of samples used for the simulations.}
  \label{tab:nGrains}
\end{table}

The attenuation curves resulting from monomodal microstructures play the role of reference in order to build a surface $\alpha(f,d)$ (\textit{i.e.} the model function) as in section \ref{sec:procedure_presentation}, while these computed in bimodal ones are used to characterize the distribution of grain size.
\subsection{Discretization}

The numerical model is based on the following system of conservation laws, composed of the balance equation of linear momentum with no source term and geometrical conservation laws:
\begin{equation}
  \label{eq:conservation_laws}
  \begin{aligned}
    & \drond{\Ucb}{t} + \sum_{i=1}^{dim}\drond{\Fcb \cdot \vect{e}_i}{x_i} = 0,\\
   & \Ucb = \bvecteur{\rho \vect{v} \\ \Cbb^{-1}:\tens{\sigma}} ; \quad \Fcb\cdot\vect{e}_i= \bvecteur{-\tens{\sigma}\cdot \vect{e}_i \\ -\frac{\vect{v}\otimes\vect{e}_i + \vect{e}_i\otimes\vect{v} }{2}}
  \end{aligned}
\end{equation}
In system \eqref{eq:conservation_laws}, the unknowns are respectively $\tens{\sigma}$ and $\vect{v}$, the Cauchy stress tensor and the velocity vector.
The solid domain presented in figure \ref{fig:problem_setting} is then discretized as a Cartesian grid of space step $\Delta x = 12.0 \mu$m, in such a way that the semi-discrete form of the above governing equations can be written by means of the discontinuous Galerkin (DG) first-order approximation \cite{Cockburn, TIE2018, TIE2020}.
Note that such a finite element size leads to good convergence properties for continuous Galerkin finite element schemes \cite{Pamel2017,Xue_noise}, which have the same (second-order) accuracy as the considered DG approach. 
At last, an explicit two-step second-order Runge-Kutta time-discretization is used to derive the discrete system.
The reader interested in more details about formulation, implementation and stability of DG methods for anisotropic and piecewise homogeneous media should refer to references \cite{TIE2018, TIE2020, TIE2019}.

The microstructural morphology is managed by overlapping \textsc{Neper} rasterized tessellations to the Cartesian grid so that each finite element can easily be identified as belonging to a grain.
Then, the elastic tensor can be computed grain-wise given Euler angles.
Recall that for one morphology (\textit{i.e. one distribution of the equivalent diameter}), several spatial distributions of the cubic axes orientation are generated following \eqref{eq:euler_angles}, which allows considering a sufficient amount of grains \review{so that the application of the analytical models is appropriate}.
The number of samples considered in each case is reported in the last column of table \ref{tab:nGrains}.
Attenuation coefficients, whose computation is explained below, are at last averaged over the samples.

\subsection{Post-treatment of the ultrasonic data}

The velocity solution provided by the finite element procedure in time domain is recorded at $N^{\text{points}}$ lying on the free end of the sample.
Those signals are then windowed in order to extract at each point the reflection of the incident wave only.
In order to take into account the dispersion resulting form the constitutive heterogeneities, the time window is defined for one point as: $t\in \[T_{v_{max}}-T_r/2,\: T_{v_{max}}+T_r/2\]$, where $T_{v_{max}}$ is the time of maximum velocity at that point.
The average of the field computed on the free end is next transformed in the frequency domain for the homogeneous reference solid and the heterogeneous sample.
At last, the attenuation coefficient is computed as:
\begin{equation}
  \label{eq:numerical_alpha}
  \alpha(f) = \frac{1}{D}\log\(\frac{\abs{v^{Iso}(f)}}{\abs{v^{Sample}(f)}}\)
\end{equation}
where $D\equiv l$ is the propagation distance and $v$ is the discrete Fourier transform of the velocity component $v_1$.

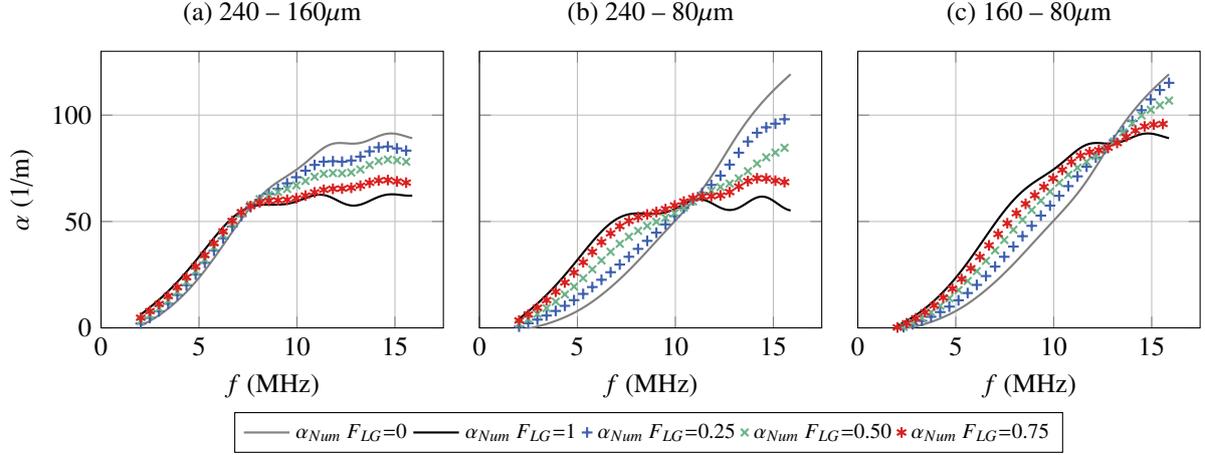
\begin{figure*}[ht]
  \centering
  \begin{tikzpicture}
  \begin{groupplot}[group style={group size=3 by 1,horizontal sep=3ex,vertical sep=0ex
      ,yticklabels at=edge left,ylabels at=edge left
    }
    ,ymin=0.,ymax=130
    ,xmin=0.
    ,width=0.37\textwidth
    ,xlabel=$f$ (MHz),ylabel=$\alpha$ (1/m)
    ]
    
    \nextgroupplot[title={(a) 240 -- 160$\mu$m}]
    
    \addplot[black!50,thick] table[x expr=\thisrow{f}/1.e6,y=FLG0] {pgfFigures/pgfFiles/bimodal240_160_2D.pgf};
    \addplot[black,thick] table[x expr=\thisrow{f}/1.e6,y=FLG100] {pgfFigures/pgfFiles/bimodal240_160_2D.pgf};


    \addplot[Blue,only marks,mark repeat=3,mark=+,thick] table[x expr=\thisrow{f}/1.e6,y=FLG25] {pgfFigures/pgfFiles/bimodal240_160_2D.pgf};
    
    \addplot[Green,only marks,mark repeat=3,mark=x,thick] table[x expr=\thisrow{f}/1.e6,y=FLG50] {pgfFigures/pgfFiles/bimodal240_160_2D.pgf};
    
    \addplot[Red,only marks,mark repeat=3,mark=asterisk,thick] table[x expr=\thisrow{f}/1.e6,y=FLG75] {pgfFigures/pgfFiles/bimodal240_160_2D.pgf};
    
    \nextgroupplot[title={(b) 240 -- 80$\mu$m}]
    
    \addplot[black!50,thick] table[x expr=\thisrow{f}/1.e6,y=FLG0] {pgfFigures/pgfFiles/bimodal160_80_2D.pgf};
    \addplot[black,thick] table[x expr=\thisrow{f}/1.e6,y=FLG100] {pgfFigures/pgfFiles/bimodal240_80_2D.pgf};


    \addplot[Blue,only marks,mark repeat=3,mark=+,thick] table[x expr=\thisrow{f}/1.e6,y=FLG25] {pgfFigures/pgfFiles/bimodal240_80_2D.pgf};
    
    \addplot[Green,only marks,mark repeat=3,mark=x,thick] table[x expr=\thisrow{f}/1.e6,y=FLG50] {pgfFigures/pgfFiles/bimodal240_80_2D.pgf};
    
    \addplot[Red,only marks,mark repeat=3,mark=asterisk,thick] table[x expr=\thisrow{f}/1.e6,y=FLG75] {pgfFigures/pgfFiles/bimodal240_80_2D.pgf};

    \nextgroupplot[title={(c) 160 -- 80$\mu$m},legend style={at={($(.5,-0.4)+(0.45cm,0.35cm)$)},legend columns=5}]
    
    \addplot[black!50,thick] table[x expr=\thisrow{f}/1.e6,y=FLG0] {pgfFigures/pgfFiles/bimodal160_80_2D.pgf};
    \addplot[black,thick] table[x expr=\thisrow{f}/1.e6,y=FLG100] {pgfFigures/pgfFiles/bimodal160_80_2D.pgf};


    \addplot[Blue,only marks,mark repeat=3,mark=+,thick] table[x expr=\thisrow{f}/1.e6,y=FLG25] {pgfFigures/pgfFiles/bimodal160_80_2D.pgf};
    
    \addplot[Green,only marks,mark repeat=3,mark=x,thick] table[x expr=\thisrow{f}/1.e6,y=FLG50] {pgfFigures/pgfFiles/bimodal160_80_2D.pgf};
    
    \addplot[Red,only marks,mark repeat=3,mark=asterisk,thick] table[x expr=\thisrow{f}/1.e6,y=FLG75] {pgfFigures/pgfFiles/bimodal160_80_2D.pgf};

    \addlegendentry{$\alpha_{Num} \: F_{LG}$=0}
    \addlegendentry{$\alpha_{Num} \: F_{LG}$=1}


    \addlegendentry{$\alpha_{Num} \: F_{LG}$=0.25}
    
    \addlegendentry{$\alpha_{Num} \: F_{LG}$=0.50}
    
    \addlegendentry{$\alpha_{Num} \: F_{LG}$=0.75}

  \end{groupplot}
\end{tikzpicture}

  \caption{Attenuation coefficient for several two-dimensional bimodal microstructures depending on the volume fraction of large grains and the modal equivalent diameters.}
  \label{fig:alpha_bimodal2D}
\end{figure*}

Figure \ref{fig:alpha_bimodal2D} shows the attenuation curves computed for nine of the two-dimensional bimodal microstructures considered so far.
As before, the results are gathered by modal diameters so that the comparison with the associated monomodal distributions is straightforward.
First, in the considered frequency range, the attenuation curves resulting from the simulation in monomodal microstructures exhibit similar shape as analytical ones (see the first row in figure \ref{fig:alphaTH_bimodal}).
Namely, the attenuation coefficient for the equivalent diameter $d=80\mu$m increases much more than the two others which almost tend to a plateau for $f=15$MHz.
However, numerical results lead to less smooth curves than analytical ones, which is most likely due to multiple-scattering.
Second, the dependence of the attenuation coefficient on the volume fraction of large grains for bimodal microstructures is well observed.
Indeed, the higher $F_{LG}$, the closer the results are to the monomodal curve associated to the large grains.
\review{Third, the comparison between Figures \ref{fig:alphaTH_bimodal} and \ref{fig:alpha_bimodal2D} shows that the crossing point of the attenuation curves occurs at a slightly different frequency in the analytical and numerical results.
  One possible explanation for this difference could be the influence of the multiple scattering captured in numerical simulations but ignored by analytical models.
  Finding a correlation between the gap in the crossing point frequency and the degree of multiple scattering would be an interesting point to check in future work.}
At last, for each combination of modal diameters, all the attenuation curves cross at the same point. 
Therefore, the numerical study presented here agrees with the analysis performed on the theoretical attenuation coefficient for bimodal microstructures.
It is then proposed to apply the procedure described in section \ref{sec:procedure_presentation} to these numerical results in the following.

\subsection{Characterization of the microstructures by inversion} 
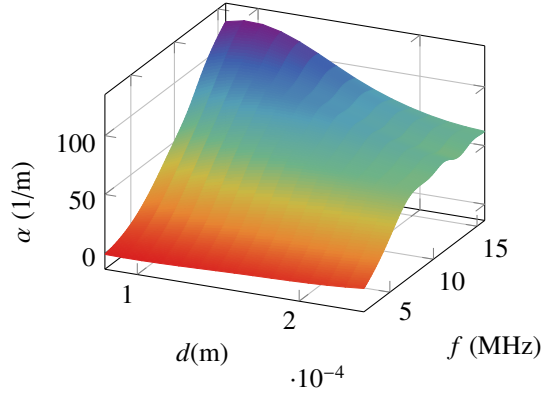
\begin{figure}[!ht]
  \centering
  \begin{tikzpicture}
  \begin{axis}[xlabel=$d $(m),ylabel=$f$ (MHz),zlabel=$\alpha$ (1/m)
    ,width=0.4\linewidth
    ]
    \addplot3[surf,shader=flat] table[x=d,y expr=\thisrow{f}/1.e6,z=alpha] {pgfFigures/pgfFiles/alpha_numericalSurface2D.pgf};
    
  \end{axis}
\end{tikzpicture}
  \caption{Reconstruction of the attenuation coefficient surface for monomodal microstructures in the ($d$,$f$,$\alpha$) space for two-dimensional cases.}
  \label{fig:numericalBimodal2D}
\end{figure}

Given that numerical and analytical attenuation coefficients show some amplitude mismatch, it is proposed to base the model function of optimization problems \eqref{eq:optimization_problem} and \eqref{eq:optimization_problem_mono} upon numerical monomodal results.
Thus, the numerical data computed for monomodal microstructures with equivalent diameter $d=\{80,120,160,200,240\}\mu$m play now the role of the reference solutions.
The bivariate cubic spline reconstruction of the attenuation surface $\alpha(f,d)$ can be seen in figure \ref{fig:numericalBimodal2D}.

As in section \ref{sec:procedure_presentation}, both monomodal and bimodal fitting are carried out for the attenuation data resulting from numerical simulations.
The optimal parameters enabling the characterization of the corresponding distribution of the grain equivalent diameters as well as the errors between the numerical and optimized curves are presented in table \ref{tab:diameqTH_bimod_num}.
\begin{table}[!ht]
  \centering

\begin{tabular}{ccc|cc|cccc}
  \multicolumn{3}{c|}{Input parameters} & \multicolumn{2}{c|}{Monomodal fitting} & \multicolumn{4}{c}{Bimodal fitting}\\
  $F_{LG}$ (\%)& $d^{SG}$ & $d^{LG}$ & $d_{opt}$ & $\epsilon_{Mono}$ (\%) &  $F^{opt}_{LG}$ (\%)& $d_{opt}^{SG}$ & $d^{LG}_{opt}$& $\epsilon_{Bi}$ (\%) \\
  \hline
  25 & 160 & 240 & 176 & 2.7 & 50 & 148 & 207 & $1.2$\\
  50 & 160 & 240 & 192 & 3.4 & 61 & 149 & 224 & $1.4$\\
  75 & 160 & 240 & 217 & 3.1 & 90 & 117 & 228 & $1.8$\\
  
  \hline
  25 & 80 & 240 & 80 & 14 & 28 & 80 & 240 & $2.3$\\
  50 & 80 & 240 & 205 & 20 & 57 & 80 & 240 & $2.9$\\
  75 & 80 & 240 & 222 & 8.9 & 82 & 80 & 236 & $2.7$\\

  \hline
  25 & 80 & 160 & 89 & 6.6 & 32 & 80 & 148 & $1.7$\\
  50 & 80 & 160 & 126 & 10  & 41 & 91 & 175 & $1.4$\\
  75 & 80 & 160 & 151 & 6.1  & 78 & 80 & 161 & $1.2$\\

  \hline
  35 & 90 & 190 & 126 & 11 & 49 & 83 & 165 & $1.1$
\end{tabular}


  \caption{Solutions of optimization problems \eqref{eq:optimization_problem_mono} and \eqref{eq:optimization_problem} applied to the two-dimensional numerical attenuation data for bimodal distributions of the equivalent diameter.}
  \label{tab:diameqTH_bimod_num}
\end{table}
As one can see, the monomodal errors are similar to those computed for the analytical data in table \ref{tab:diameqTH_bimod2D} whereas bimodal errors are greater in that case.
This means that the optimization procedure struggles to fit parameters with the numerical bimodal attenuation data.
In addition, the comparison between monomodal and bimodal errors shows that $\epsilon_{Bi}<\epsilon_{Mono}$ for all the considered distributions of equivalent diameter.
Therefore, in case we do not know \textit{a priori} the microstructure, a high monomodal error can indicate that the distribution of the grain size is actually bimodal.
Then, both errors take maximal values for a high ratio $d^{LG}/d^{SG}$, which has been already observed in section \ref{sec:procedure_presentation}.
In addition, the results of the bimodal optimization for the combination $d^{LG}=240 \mu m$, $d^{SG}=80\mu$m must be taken carefully.
Indeed, those values of the equivalent diameter are also the bounds of the definition domain of $\alpha(f,d)$.
It is therefore probable that these solutions are found by lack of other values $\alpha(f,d<80\mu m)$ and $\alpha(f,d>240\mu m)$, which would explain the high amount of error while the set of optimized parameters are in that case very close from the actual one.

\begin{figure}[ht]
  \centering
  \begin{tikzpicture}
  \begin{axis}[
    ,width=0.50\textwidth
    ,xlabel=$f$ (MHz),ylabel=$\alpha$ (1/m)
    ,legend pos=south east
    ]

    \addplot[black,only marks,mark repeat=3,mark=+,very thick] table[x expr=\thisrow{f}/1.e6,y=FLG35] {pgfFigures/pgfFiles/bimodal190_90_2D.pgf};

    \addplot[Blue,very thick] table[x expr=\thisrow{f}/1.e6,y=bi35] {pgfFigures/pgfFiles/fitResult2D_190_90mm.pgf};

    \addplot[Red,densely dotted,very thick] table[x expr=\thisrow{f}/1.e6,y=mono35] {pgfFigures/pgfFiles/fitResult2D_190_90mm.pgf};

    \addlegendentry{Numerical data }
    
    \addlegendentry{Bimodal fitting}
    
    \addlegendentry{Monomodal fitting}

  \end{axis}
\end{tikzpicture}

  \caption{Comparison between numerical data and optimized curves for the bimodal distribution of the equivalent diameters with modal values $d=\{90,190\}\mu m$ and volume fraction of large grains $F_{LG}=0.35$.}
  \label{fig:2Dfit_190_90}
\end{figure}
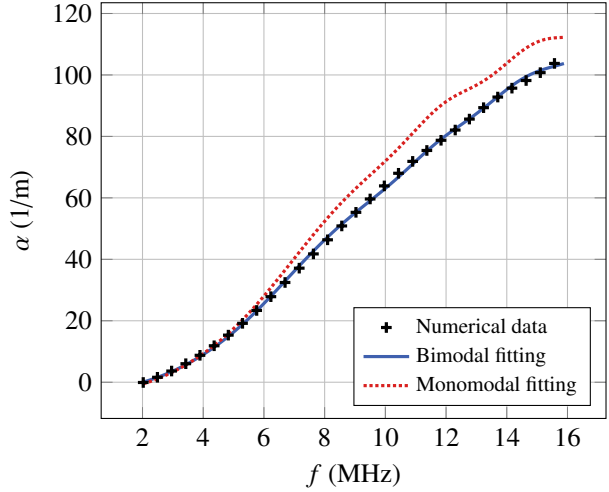

We now focus on the last row of table \ref{tab:diameqTH_bimod_num}, which concerns a bimodal distribution for which none of the monomodal attenuation curves is known.
One sees that the characterization of the distribution of equivalent diameter is very good, even though the volume fraction of large grain is slightly overestimated.
Figure \ref{fig:2Dfit_190_90} shows a comparison of the numerical data and the optimized curves for this microstructure.
Those curves highlight that the monomodal fitting is close to the data for low frequencies only whereas the bimodal fitting agrees well in overall. 
This first emphasizes the importance of considering a large frequency range for the optimization.
Indeed, a monomodal solution can be found if only a small portion of the attenuation curve is looked at, even though the distribution is really bimodal.
Second, the gaps between the monomodal fitting and the numerical data indicate that a bimodal distribution should be considered, namely, better results should be computed with the bimodal fitting.
%
Doing so, the model function can accurately be fitted, which results in morphological parameters that describe the microstructure satisfactorily.
However, it is believed that increase the precision on the $\alpha(f,d)$ surface reconstruction would enhance the optimization as mentioned in the previous section.

The results presented here pave the way for the use of the bimodal fitting in order to identify and characterize bimodal distributions of the grain size in polycrystals.
Indeed, the approach followed here for visualization purposes has been applied to attenuation data provided by numerical simulations but could easily be extended to experimental data.
The key point of the method is that it requires a reliable model function which, in the present work, is reconstructed from approximate solutions computed in monomodal microstructures, which may lack of accuracy. 
However, it is believed that the experimenter having a precise attenuation database for monomodal microstructures is able to accurately characterize bimodal ones.





\section{Concluding remarks}
\label{sec:conclusion}
In this paper, the influence of bimodal distributions of the equivalent diameter in polycrystals has been investigated.
First, the spatial correlation function $W(r)$ in polycrystalline materials made of equiaxed grains has been shown to yield an additive decomposition of the attenuation coefficient in section \ref{sec:theory}.
More specifically, the volume fraction of large grains plays an important role in the analytical expression of the frequency-dependent attenuation coefficient.
This property has then been confirmed in section \ref{sec:procedure} owing to the joint developments of a procedure to numerically model bimodal microstructures with the software \textsc{Neper} and of analytical attenuation formulas in two and three space dimensions taking into account a generic spatial correlation function.
From these results, a \textit{least squares minimization-based} numerical procedure for characterizing bimodal distributions of the grains' equivalent diameter has been presented.
The aforementioned approach has been shown to yield very good results for analytical attenuation curves in both two and three space dimensions as well as two-dimensional numerical attenuation data.
Namely, for the considered bimodal microstructures, the volume fraction of large grains and the modal equivalent diameters of both large and small grains have been satisfyingly approximated by applying the proposed inversion procedure to attenuation curves.

The presented characterization method, which has been illustrated for analytical and numerical attenuation data, could be applied to experimental ones.
In that context, the model function's surface $\alpha(f,d)$ could be built using either experimental or numerical attenuation curves computed for monomodal distributions of the grain size.
A numerical--experimental mixed procedure however requires the use of numerical approaches accurate enough to guarantee a good agreement between simulation and testing.
On the other hand, the present study should be supplemented with an analysis of the effect of the distributions width.
Indeed, the same standard deviation has been considered for both large grains and small grains in bimodal distributions, namely $\sigma^{LG}=\sigma^{SG}$.
Nevertheless, it has been shown that the distribution width has a measurable effect on attenuation in the Rayleigh and transition regions \cite{Arguelles2017}, in such a way that variations of $\sigma^{LG}$ and $\sigma^{SG}$ should have an impact on the optimization procedure.
Moreover, the attenuation series \eqref{eq:multiphase_decomposition} suggest that a grain size distribution can be discretized so that a multi-parameter optimization can be done, which is similar to the approach presented in \cite{Nicoletti,nicoletti1997}.
\textit{Thick} grain size distributions could thus be considered by using \textit{thin} monomodal distributions, namely, with a small deviation, to build the model function.
For application to the characterization of actual microstructures, the numerical modeling of \textit{thin} distributions should highly increase the flexibility of the approach, which once again strengthen the development of a mixed procedure.
This is however, a more long term perspective of the present research.



\section*{Acknowledgments}
This work was granted by the 9th research program of Labex LaSIPS (Paris-Saclay Systems and Engineering Laboratory, Paris-Saclay University). Computations
were performed using HPC resources from the computing centre of CentraleSup{\'e}lec and ENS Paris-Saclay.

\appendix{}
\onecolumn
\section{Semi-analytical model}
\label{sec:alpha_formula}
The equation \eqref{eq:alpha_integral} is developed in detail hereafter.

Let us first rewrite the equation for both two and three dimensional cases:
\begin{equation}
  \label{eq:alpha_integral2D_3D}
  \alpha^{\beta \gamma} =\left\lbrace
  \begin{aligned}
    & \sum_{k,l,m,n}\text{Im}\[ \frac{k_{0\beta}\left\langle \delta C_{J^\beta J kl} \delta C_{mn J^\beta J} \right\rangle}{2C^0_{J^\beta J J^\beta J}} \int\displaylimits_{r=0}^\infty \int\displaylimits_{\theta=0}^{2\pi} G^\gamma_{km}(r,\theta) D^{\beta,J}_{ln}(r,\theta) r dr d\theta \] \quad \text{in 2D}  \\
    & \sum_{k,l,m,n}\text{Im}\[ \frac{k_{0\beta}\left\langle \delta C_{J^\beta J kl} \delta C_{mn J^\beta J} \right\rangle}{2C^0_{J^\beta J J^\beta J}} \int\displaylimits_{r=0}^\infty \int\displaylimits_{\theta=0}^{2\pi} \int\displaylimits_{\varphi=0}^{\pi} G^\gamma_{km}(r,\theta,\varphi) D^{\beta,J}_{ln}(r,\theta,\varphi) r^2 \sin \varphi dr d\theta d\varphi\]  \quad \text{in 3D}
  \end{aligned}
  \right.
\end{equation}
in which the expressions of the dyadic Green\review{'s} function tensor $\tens{G}^\gamma$ and $\tens{D}^{\beta,J}$ are:
\begin{align}
  \label{eq:Ggamma_expression_appendix}
  & \tens{G}^\gamma(\vect{r})= \frac{1-c(\gamma)}{4\pi \rho \omega^2}\( A_{rr}(k_{0\gamma}r)\frac{\vect{r}\otimes\vect{r}}{r^2} -A_{I}(k_{0\gamma}r)\tens{I}\)\\
  \label{eq:Dbeta_expression_appendix}
  &\tens{D}^{\beta,J}(\vect{r})= \nablat_{\vect{r}}\(\nablat_{\vect{r}}\(W(r) e^{ik_{0\beta}\vect{e}_{J} \cdot \vect{r}}\)\)
\end{align}

Depending on the space dimension, the expressions of functions $A_{rr}(\bullet)$ and $A_I(\bullet)$ read: 
\begin{align}
  \label{eq:Aij_components}
  & A^{3D}_{rr}(k_{0\gamma}r) = \frac{e^{ik_{0\gamma}r}}{r^3}\[3-3ik_{0\gamma}r- k_{0\gamma}^2r^2 \]\\
  & A^{3D}_{I}(k_{0\gamma}r)= \frac{e^{ik_{0\gamma}r}}{r^3}\[1-ik_{0\gamma}r- \frac{c(\gamma)}{2}k_{0\gamma}^2r^2 \]\\
  & A^{2D}_{rr}(k_{0\gamma}r)=i\pi\[ \frac{k_{0\gamma}}{r}H^{(1)}_1(k_{0\gamma}r) - \frac{k^2_{0\gamma}}{2}\( H^{(1)}_0(k_{0\gamma}r) -H^{(1)}_2(k_{0\gamma}r)\)\] \\
  & A^{2D}_{I}(k_{0\gamma}r)=i\pi\[ \frac{k_{0\gamma}}{r}H^{(1)}_1(k_{0\gamma}r) - c(\gamma)\frac{k^2_{0\gamma}}{2} H^{(1)}_0(k_{0\gamma}r) \]
\end{align}
where $H^{(1)}_q(\bullet)$ denotes the Hankel function of first kind.

The multiple integrals in equation \eqref{eq:alpha_integral2D_3D} can be reduced to a single integration over the radius, which can highly reduce the computational cost of numerical integration techniques. 
This is first done for two-dimensional problems by expending the integrands according to equation \eqref{eq:generic_D}  for $k,l,m,n={P,Q}$ such that $\vect{e}_P$ is the propagation direction and $\vect{e}_Q$ is perpendicular to it.
Then, the integration over $\theta$ can be performed.
Those integrals, which involve Bessel functions of the first kind, are denoted as $\hbar^{2D}_K(k_{0\beta}r)$ in what follows (see \cite{Xue_2Dalpha} for further details).
In two space dimensions, the notation $ \int\displaylimits_{\theta=0}^{2\pi} G^\gamma_{km}(r,\theta) D^{\beta,J}_{ln}(r,\theta) d\theta  = \eta^{\gamma\beta,J}_{kmln}(r)$, so that $\alpha^{\beta \gamma} =\sum_{k,l,m,n}\text{Im}\[ \frac{k_{0\beta}\left\langle \delta C_{J^\beta J kl} \delta C_{mn J^\beta J} \right\rangle}{2C^0_{J^\beta J J^\beta J}} \int\displaylimits_{r=0}^\infty \eta^{\beta\gamma,J}_{kmln}(r) r dr \]$ , leads to the derivation of the following five non-zero terms: 
\begin{align}
    &\begin{aligned} 
    \frac{4\pi\rho\omega^2}{1-c(\gamma)}\eta^{\gamma\beta,P}_{PPPP}(r)= A_{rr}(k_{0\gamma}r) &\[\(W''-\frac{W'}{r}\)\hbar^{2D}_{C4}(k_{0\beta}r) + \(\frac{W'}{r}- k^2_{0\beta}W\)\hbar^{2D}_{C2}(k_{0\beta}r) +2ik_{0\beta}W'\hbar^{2D}_{C3}(k_{0\beta}r) \] \\
    -A_I(k_{0\gamma}r)&\[ \(W''-\frac{W'}{r}\)\hbar^{2D}_{C2}(k_{0\beta}r) + \(\frac{W'}{r}- k^2_{0\beta}W\)\hbar^{2D}_{S0}(k_{0\beta}r) +2ik_{0\beta}W'\hbar^{2D}_{C1}(k_{0\beta}r)\]
  \end{aligned}\\
  &\begin{aligned}
  \frac{4\pi\rho\omega^2}{1-c(\gamma)}\eta^{\gamma\beta,P}_{QQQQ}(r) = A_{rr} (k_{0\gamma}r)&\[\(W''-\frac{W'}{r}\)\hbar^{2D}_{S4}(k_{0\beta}r)+ \frac{W'}{r}\hbar^{2D}_{S2}(k_{0\beta}r) \] \\
    -A_I(k_{0\gamma}r)&\[ \(W''-\frac{W'}{r}\)\hbar^{2D}_{S2}(k_{0\beta}r) + \frac{W'}{r}\hbar^{2D}_{S0}(k_{0\beta}r)\]
  \end{aligned}\\
  &\begin{aligned} 
    \frac{4\pi\rho\omega^2}{1-c(\gamma)}\eta^{\gamma\beta,P}_{PPQQ}(r) = A_{rr}(k_{0\gamma}r) &\[\(W''-\frac{W'}{r}\)\hbar^{2D}_{S2C2}(k_{0\beta}r) + \frac{W'}{r}\hbar^{2D}_{C2}(k_{0\beta}r)  \] \\
    -A_I(k_{0\gamma}r)&\[ \(W''-\frac{W'}{r}\)\hbar^{2D}_{S2}(k_{0\beta}r) + \frac{W'}{r}\hbar^{2D}_{S0}(k_{0\beta}r)\]
  \end{aligned}\\
  &\begin{aligned} 
    \frac{4\pi\rho\omega^2}{1-c(\gamma)}\eta^{\gamma\beta,P}_{QQPP}(r) = A_{rr} (k_{0\gamma}r)&\[\(W''-\frac{W'}{r}\)\hbar^{2D}_{S2C2}(k_{0\beta}r)+ \(\frac{W'}{r}- k^2_{0\beta}W\)\hbar^{2D}_{S2}(k_{0\beta}r)+2ik_{0\beta}W'\hbar^{2D}_{S2C1}(k_{0\beta}r) \] \\
    -A_I(k_{0\gamma}r)&\[ \(W''-\frac{W'}{r}\)\hbar^{2D}_{C2}(k_{0\beta}r) + \(\frac{W'}{r}- k^2_{0\beta}W\)\hbar^{2D}_{S0}(k_{0\beta}r) +2ik_{0\beta}W'\hbar^{2D}_{C1}(k_{0\beta}r)\]
  \end{aligned}\\
  & \frac{4\pi\rho\omega^2}{1-c(\gamma)}\eta^{\gamma\beta,P}_{PQPQ}(r) = A_{rr} (k_{0\gamma}r)\[\(W''-\frac{W'}{r}\)\hbar^{2D}_{S2C2} (k_{0\beta}r)+ ik_{0\beta}W'\hbar^{2D}_{S2C1}(k_{0\beta}r) \]
\end{align}

For three-dimensional cases, two additional terms have to be considered.
In that case, one writes $\eta^{\gamma\beta,J}_{kmln}(r) = \int\displaylimits_{\theta=0}^{2\pi}\int\displaylimits_{\varphi=0}^{\pi} G^\gamma_{km}(r,\theta,\varphi) D^{\beta,J}_{ln}(r,\theta,\varphi) \sin \varphi  d\varphi d\theta$, in such a way that $\alpha^{\beta \gamma} =\sum_{k,l,m,n}\text{Im}\[ \frac{k_{0\beta}\left\langle \delta C_{J^\beta J kl} \delta C_{mn J^\beta J} \right\rangle}{2C^0_{J^\beta J J^\beta J}} \int\displaylimits_{r=0}^\infty \eta^{\gamma\beta,J}_{kmln}(r) r^2 dr\]$, and:
\begin{align}
  &\begin{aligned} 
    \frac{4\pi\rho\omega^2}{1-c(\gamma)}\eta^{\gamma\beta,P}_{PPPP}(r) = 2\pi A_{rr} (k_{0\gamma}r)&\[\(W''-\frac{W'}{r}\)\hbar^{3D}_{C4S1}(k_{0\beta}r) + \(\frac{W'}{r}- k^2_{0\beta}W\)\hbar^{3D}_{C2S1}(k_{0\beta}r) +2ik_{0\beta}W'\hbar^{3D}_{C3S1}(k_{0\beta}r) \] \\
    -2\pi A_I(k_{0\gamma}r)&\[ \(W''-\frac{W'}{r}\)\hbar^{3D}_{C2S1}(k_{0\beta}r) + \(\frac{W'}{r}- k^2_{0\beta}W\)\hbar^{3D}_{S1} (k_{0\beta}r)+2ik_{0\beta}W'\hbar^{3D}_{C1S1}(k_{0\beta}r)\]
  \end{aligned}\\
  &\begin{aligned}
  \frac{4\pi\rho\omega^2}{1-c(\gamma)}\eta^{\gamma\beta,P}_{QQQQ}(r) = \pi A_{rr}(k_{0\gamma}r) &\[\frac{3}{4}\(W''-\frac{W'}{r}\)\hbar^{3D}_{S5}(k_{0\beta}r)+ \frac{W'}{r}\hbar^{3D}_{S3}(k_{0\beta}r) \] \\
    -\pi A_I(k_{0\gamma}r)&\[ \(W''-\frac{W'}{r}\)\hbar^{3D}_{S3}(k_{0\beta}r) + 2 \frac{W'}{r}\hbar^{3D}_{S1}(k_{0\beta}r)\]
  \end{aligned}\\
  &\begin{aligned} 
    \frac{4\pi\rho\omega^2}{1-c(\gamma)}\eta^{\gamma\beta,P}_{PPQQ}(r) = \pi A_{rr}(k_{0\gamma}r) &\[\(W''-\frac{W'}{r}\)\hbar^{3D}_{S3C2}(k_{0\beta}r) + 2\frac{W'}{r}\hbar^{3D}_{S1C2}(k_{0\beta}r)  \] \\
    -\pi A_I(k_{0\gamma}r)&\[ \(W''-\frac{W'}{r}\)\hbar^{3D}_{S3}(k_{0\beta}r) + 2\frac{W'}{r}\hbar^{3D}_{S1}(k_{0\beta}r)\]
  \end{aligned}\\
  &\begin{aligned} 
    \frac{4\pi\rho\omega^2}{1-c(\gamma)}\eta^{\gamma\beta,P}_{QQPP}(r) = \pi A_{rr} (k_{0\gamma}r)&\[\(W''-\frac{W'}{r}\)\hbar^{3D}_{S3C2}(k_{0\beta}r)+ \(\frac{W'}{r}- k^2_{0\beta}W\)\hbar^{3D}_{S3}(k_{0\beta}r)+2ik_{0\beta}W'\hbar^{3D}_{S3C1}(k_{0\beta}r) \] \\
    -2\pi A_I(k_{0\gamma}r)&\[ \(W''-\frac{W'}{r}\)\hbar^{3D}_{S1C2}(k_{0\beta}r) +\(\frac{W'}{r}- k^2_{0\beta}W\)\hbar^{3D}_{S1}(k_{0\beta}r) +2ik_{0\beta}W'\hbar^{3D}_{S1C1}(k_{0\beta}r)\]
  \end{aligned}\\
  & \frac{4\pi\rho\omega^2}{1-c(\gamma)}\eta^{\gamma\beta,P}_{PQPQ}(r) = \pi A_{rr}(k_{0\gamma}r) \[\(W''-\frac{W'}{r}\)\hbar^{3D}_{S3C2}(k_{0\beta}r) + ik_{0\beta}W'\hbar^{3D}_{S3C1}(k_{0\beta}r) \]\\
  &\begin{aligned} 
    \frac{4\pi\rho\omega^2}{1-c(\gamma)}\eta^{\gamma\beta,P}_{QQKK}(r) = \pi A_{rr} (k_{0\gamma}r)&\[\frac{1}{4}\(W''-\frac{W'}{r}\)\hbar^{3D}_{S5}(k_{0\beta}r) + \frac{W'}{r}\hbar^{3D}_{S3}(k_{0\beta}r)  \] \\
    -\pi A_I(k_{0\gamma}r)&\[ \(W''-\frac{W'}{r}\)\hbar^{3D}_{S3}(k_{0\beta}r) + 2\frac{W'}{r}\hbar^{3D}_{S1}(k_{0\beta}r)\]
  \end{aligned}\\
  & \frac{4\pi\rho\omega^2}{1-c(\gamma)}\eta^{\gamma\beta,P}_{QKQK}(r) = \frac{\pi}{4}A_{rr}(k_{0\gamma}r) \(W''-\frac{W'}{r}\)\hbar^{3D}_{S5}(k_{0\beta}r)  
\end{align}
Analogously to two-dimensional cases, the functions $\hbar^{3D}_K(k_{0\beta}r)$ result from the integration over $\varphi$ and their expressions can be found in \cite{Xue_2Dalpha}, and the integration over $\theta$ results in the multiplier $2\pi$.

\section*{References}
\bibliographystyle{elsarticle-num}
\bibliography{Biblio}

\begin{thebibliography}{10}
\expandafter\ifx\csname url\endcsname\relax
  \def\url#1{\texttt{#1}}\fi
\expandafter\ifx\csname urlprefix\endcsname\relax\def\urlprefix{URL }\fi
\expandafter\ifx\csname href\endcsname\relax
  \def\href#1#2{#2} \def\path#1{#1}\fi

\bibitem{smith}
R.~Smith, The effect of grain size distribution on the frequency dependence of
  the ultrasonic attenuation in polycrystalline materials, Ultrasonics 20~(5)
  (1982) 211--214.

\bibitem{Nicoletti}
D.~Nicoletti, N.~Bilgutay, B.~Onaral, Power‐law relationships between the
  dependence of ultrasonic attenuation on wavelength and the grain size
  distribution, The Journal of the Acoustical Society of America 91~(6) (1992)
  3278--3284.
\newblock \href {http://dx.doi.org/10.1121/1.402862}
  {\path{doi:10.1121/1.402862}}.

\bibitem{botvina}
L.~Botvina, L.~J. Fradkin, B.~Bridge, A new method for assessing the mean grain
  size of polycrystalline materials using ultrasonic nde, Journal of materials
  science 35~(18) (2000) 4673--4683.

\bibitem{PALANICHAMY}
P.~Palanichamy, A.~Joseph, T.~Jayakumar, B.~Raj, Ultrasonic velocity
  measurements for estimation of grain size in austenitic stainless steel, NDT
  \& E International 28~(3) (1995) 179 -- 185.
\newblock \href
  {http://dx.doi.org/https://doi.org/10.1016/0963-8695(95)00011-L}
  {\path{doi:https://doi.org/10.1016/0963-8695(95)00011-L}}.

\bibitem{sundin}
S.~Sundin, D.~Artymowicz, Direct measurements of grain size in low-carbon
  steels using the laser ultrasonic technique, Metallurgical and Materials
  Transactions A 33~(3) (2002) 687--691.

\bibitem{bouda}
A.~B. Bouda, S.~Lebaili, A.~Benchaala, Grain size influence on ultrasonic
  velocities and attenuation, {Ndt \& E International} 36~(1) (2003) 1--5.

\bibitem{li}
X.~Li, Y.~Song, F.~Liu, H.~Hu, P.~Ni, Evaluation of mean grain size using the
  multi-scale ultrasonic attenuation coefficient, NDT \& E International 72
  (2015) 25--32.

\bibitem{levesque}
D.~L{\'e}vesque, S.~Kruger, G.~Lamouche, R.~Kolarik~II, G.~Jeskey, M.~Choquet,
  J.-P. Monchalin, Thickness and grain size monitoring in seamless tube-making
  process using laser ultrasonics, NDT \& E International 39~(8) (2006)
  622--626.

\bibitem{sarkar}
S.~Sarkar, A.~Moreau, M.~Militzer, W.~Poole, Evolution of austenite
  recrystallization and grain growth using laser ultrasonics, Metallurgical and
  Materials Transactions A 39~(4) (2008) 897--907.

\bibitem{garcin}
T.~Garcin, J.~H. Schmitt, M.~Militzer, \textit{In-situ} laser ultrasonic grain
  size measurement in superalloy inconel 718, Journal of Alloys and Compounds
  670 (2016) 329--336.

\bibitem{keyvani}
M.~Keyvani, T.~Garcin, D.~Fabr{\`e}gue, M.~Militzer, K.~Yamanaka, A.~Chiba,
  Continuous measurements of recrystallization and grain growth in cobalt super
  alloys, Metallurgical and Materials Transactions A 48~(5) (2017) 2363--2374.

\bibitem{Stanke_Kino84}
F.~E. Stanke, G.~S. Kino, A unified theory for elastic wave propagation in
  polycrystalline materials, The Journal of the Acoustical Society of America
  75~(3) (1984) 665--681.
\newblock \href {http://dx.doi.org/10.1121/1.390577}
  {\path{doi:10.1121/1.390577}}.

\bibitem{Keller64}
F.~C. Karal, J.~B. Keller, Elastic, electromagnetic, and other waves in a
  random medium, Journal of Mathematical Physics 5~(4) (1964) 537--547.
\newblock \href {http://dx.doi.org/10.1063/1.1704145}
  {\path{doi:10.1063/1.1704145}}.

\bibitem{Weaver_modeConversion}
R.~Weaver, Diffusivity of ultrasound in polycrystals, Journal of the Mechanics
  and Physics of Solids 38~(1) (1990) 55 -- 86.
\newblock \href
  {http://dx.doi.org/https://doi.org/10.1016/0022-5096(90)90021-U}
  {\path{doi:https://doi.org/10.1016/0022-5096(90)90021-U}}.

\bibitem{Kube_JASA2017}
C.~M. Kube, Iterative solution to bulk wave propagation in polycrystalline
  materials., Journal of the Acoustical Society of America 141~(3) (2017)
  1804--1811.

\bibitem{Xue_2Dalpha}
X.~Bai, B.~Tie, J.-H. Schmitt, D.~Aubry, Comparison of ultrasonic attenuation
  within two- and three-dimensional polycrystalline media, Ultrasonics 100
  (2020) 105980.
\newblock \href
  {http://dx.doi.org/https://doi.org/10.1016/j.ultras.2019.105980}
  {\path{doi:https://doi.org/10.1016/j.ultras.2019.105980}}.

\bibitem{Arguelles2017}
A.~P. Arguelles, J.~A. Turner, Ultrasonic attenuation of polycrystalline
  materials with a distribution of grain sizes, The Journal of the Acoustical
  Society of America 141~(6) (2017) 4347--4353.
\newblock \href {http://dx.doi.org/10.1121/1.4984290}
  {\path{doi:10.1121/1.4984290}}.

\bibitem{nicoletti1997}
D.~Nicoletti, A.~Anderson, Determination of grain-size distribution from
  ultrasonic attenuation: transformation and inversion, The Journal of the
  Acoustical Society of America 101~(2) (1997) 686--689.

\bibitem{Boyce2015_bimodal}
B.~L. Boyce, T.~A. Furnish, H.~Padilla, D.~Van~Campen, A.~Mehta, Detecting
  rare, abnormally large grains by {X}-ray diffraction, Journal of materials
  science 50~(20) (2015) 6719--6729.

\bibitem{Liu2008}
D.~Liu, J.~A. Turner, Influence of spatial correlation function on attenuation
  of ultrasonic waves in two-phase materials, The Journal of the Acoustical
  Society of America 123~(5) (2008) 2570--2576.
\newblock \href {http://dx.doi.org/10.1121/1.2896757}
  {\path{doi:10.1121/1.2896757}}.

\bibitem{Ryzy2018}
M.~Ryzy, T.~Grabec, P.~Sedl{\'{a}}k, I.~A. Veres, Influence of grain morphology
  on ultrasonic wave attenuation in polycrystalline media with statistically
  equiaxed grains, The Journal of the Acoustical Society of America 143~(1)
  (2018) 219--229.
\newblock \href {http://dx.doi.org/10.1121/1.5020785}
  {\path{doi:10.1121/1.5020785}}.

\bibitem{Pamel2018}
A.~V. Pamel, G.~Sha, M.~J.~S. Lowe, S.~I. Rokhlin, Numerical and analytic
  modelling of elastodynamic scattering within polycrystalline materials, The
  Journal of the Acoustical Society of America 143~(4) (2018) 2394--2408.
\newblock \href {http://dx.doi.org/10.1121/1.5031008}
  {\path{doi:10.1121/1.5031008}}.

\bibitem{Chakrabarti2009_bimodal}
D.~Chakrabarti, M.~Strangwood, C.~Davis, Effect of bimodal grain size
  distribution on scatter in toughness, Metallurgical and Materials
  Transactions A 40~(4) (2009) 780--795.

\bibitem{Sabzi2016_bimodal}
H.~E. Sabzi, A.~{Zarei Hanzaki}, H.~Abedi, R.~Soltani, A.~Mateo, J.~Roa, The
  effects of bimodal grain size distributions on the work hardening behavior of
  a transformation-twinning induced plasticity steel, Materials Science and
  Engineering: A 678 (2016) 23 -- 32.
\newblock \href {http://dx.doi.org/https://doi.org/10.1016/j.msea.2016.09.085}
  {\path{doi:https://doi.org/10.1016/j.msea.2016.09.085}}.

\bibitem{Azizi2007_bimodal_biphase}
H.~Azizi-Alizamini, M.~Militzer, W.~Poole, A novel technique for developing
  bimodal grain size distributions in low carbon steels, Scripta Materialia
  57~(12) (2007) 1065 -- 1068.
\newblock \href
  {http://dx.doi.org/https://doi.org/10.1016/j.scriptamat.2007.08.035}
  {\path{doi:https://doi.org/10.1016/j.scriptamat.2007.08.035}}.

\bibitem{Mahesh2012bimodal}
B.~Mahesh, R.~S. Raman, C.~C. Koch, Bimodal grain size distribution: an
  effective approach for improving the mechanical and corrosion properties of
  fe--cr--ni alloys, Journal of Materials Science 47~(22) (2012) 7735--7743.

\bibitem{Zhang04}
X.-G. Zhang, W.~A. Simpson, J.~M. Vitek, D.~J. Barnard, L.~J. Tweed, J.~Foley,
  Ultrasonic attenuation due to grain boundary scattering in copper and
  copper-aluminum, The Journal of the Acoustical Society of America 116~(1)
  (2004) 109--116.
\newblock \href {http://dx.doi.org/10.1121/1.1744752}
  {\path{doi:10.1121/1.1744752}}.

\bibitem{zeng}
F.~Zeng, S.~R. Agnew, B.~Raeisinia, G.~R. Myneni, Ultrasonic attenuation due to
  grain boundary scattering in pure niobium, Journal of Nondestructive
  Evaluation 29~(2) (2010) 93--103.

\bibitem{Cockburn}
B.~Cockburn, Discontinuous {G}alerkin methods for convection-dominated
  problems, in: High-order methods for computational physics, Springer, 1999,
  pp. 69--224.

\bibitem{TIE2018}
B.~Tie, A.-S. Mouronval, V.-D. Nguyen, L.~Series, D.~Aubry, A unified
  variational framework for the space discontinuous {G}alerkin method for
  elastic wave propagation in anisotropic and piecewise homogeneous media,
  Computer Methods in Applied Mechanics and Engineering 338 (2018) 299 -- 332.
\newblock \href {http://dx.doi.org/https://doi.org/10.1016/j.cma.2018.04.018}
  {\path{doi:https://doi.org/10.1016/j.cma.2018.04.018}}.

\bibitem{Neper}
Neper Reference Manual, http://neper.sourceforge.net/ (2019).

\bibitem{Kube_WM2015}
C.~M. Kube, J.~A. Turner, Ultrasonic attenuation in polycrystals using a
  self-consistent approach., Wave Motion 57 (2015) 182--193.

\bibitem{Torquato_book}
S.~Torquato, Microstructural descriptors, in: {Random Heterogeneous Materials:
  Microstructures and Macroscopic Properties}, Interdisciplinary Applied
  Mathematics, Springer New York, 2005, pp. 23--58.

\bibitem{Roney}
R.~K. Roney, {The influence of metal grain size on the attenuation of an
  ultrasonic wave}, Ph.D. thesis, {California Institute of Technology} (1950).

\bibitem{Quey_gg}
R.~Quey, L.~Renversade, Optimal polyhedral description of 3{D} polycrystals:
  Method and application to statistical and synchrotron {X}-ray diffraction
  data, Computer Methods in Applied Mechanics and Engineering 330 (2018) 308 --
  333.
\newblock \href {http://dx.doi.org/https://doi.org/10.1016/j.cma.2017.10.029}
  {\path{doi:https://doi.org/10.1016/j.cma.2017.10.029}}.

\bibitem{Man_Wfunction}
C.-S. Man, R.~Paroni, Y.~Xiang, E.~A. Kenik, On the geometric autocorrelation
  function of polycrystalline materials, Journal of Computational and Applied
  Mathematics 190~(1) (2006) 200 -- 210, special Issue: International
  Conference on Mathematics and its Application.

\bibitem{MatParam}
W.~Petry, A.~Heiming, J.~Trampenau, M.~Alba, C.~Herzig, H.~R. Schober, G.~Vogl,
  Phonon dispersion of the bcc phase of group-{IV} metals. i. bcc titanium,
  Phys. Rev. B 43 (1991) 10933--10947.
\newblock \href {http://dx.doi.org/10.1103/PhysRevB.43.10933}
  {\path{doi:10.1103/PhysRevB.43.10933}}.

\bibitem{Sha2018}
G.~Sha, Correlation of elastic wave attenuation and scattering with volumetric
  grain size distribution for polycrystals of statistically equiaxed grains,
  Wave Motion 83 (2018) 102 -- 110.
\newblock \href
  {http://dx.doi.org/https://doi.org/10.1016/j.wavemoti.2018.08.012}
  {\path{doi:https://doi.org/10.1016/j.wavemoti.2018.08.012}}.

\bibitem{BivariateSpline}
T.~Zhou, M.-J. Lai, Scattered data interpolation by bivariate splines with
  higher approximation order, Journal of Computational and Applied Mathematics
  242 (2013) 125 -- 140.
\newblock \href {http://dx.doi.org/https://doi.org/10.1016/j.cam.2012.10.025}
  {\path{doi:https://doi.org/10.1016/j.cam.2012.10.025}}.

\bibitem{TIE2020}
B.~Tie, A.-S. Mouronval, Systematic development of upwind numerical fluxes for
  the space discontinuous {G}alerkin method applied to elastic wave propagation
  in anisotropic and heterogeneous media with physical interfaces, Computer
  Methods in Applied Mechanics and Engineering 372 (2020) 113352.
\newblock \href {http://dx.doi.org/https://doi.org/10.1016/j.cma.2020.113352}
  {\path{doi:https://doi.org/10.1016/j.cma.2020.113352}}.

\bibitem{Pamel2017}
A.~V. Pamel, G.~Sha, S.~I. Rokhlin, M.~J.~S. Lowe, Finite-element modelling of
  elastic wave propagation and scattering within heterogeneous media,
  Proceedings of the Royal Society A: Mathematical, Physical and Engineering
  Science 473~(2197) (2017) 20160738.
\newblock \href {http://dx.doi.org/10.1098/rspa.2016.0738}
  {\path{doi:10.1098/rspa.2016.0738}}.

\bibitem{Xue_noise}
X.~Bai, B.~Tie, J.-H. Schmitt, D.~Aubry, Finite element modeling of grain size
  effects on the ultrasonic microstructural noise backscattering in
  polycrystalline materials, Ultrasonics 87 (2018) 182--202.
\newblock \href {http://dx.doi.org/10.1016/j.ultras.2018.02.008}
  {\path{doi:10.1016/j.ultras.2018.02.008}}.

\bibitem{TIE2019}
B.~Tie, Some comparisons and analyses of time or space discontinuous {G}alerkin
  methods applied to elastic wave propagation in anisotropic and heterogeneous
  media, Advanced Modeling and Simulation in Engineering Sciences 6~(1) (2019)
  1--27.

\end{thebibliography}

\end{document}